\documentclass[onecolumn]{aastex63}
\usepackage{tabularx}
\usepackage{amsmath}
\usepackage[bottom]{footmisc}
\usepackage{rotating}
\usepackage[referable]{threeparttablex}
\usepackage{supertabular}
\usepackage{xspace}

\usepackage{float}

\received{10 Aug, 2021}
\revised{25 Jan, 2022}
\accepted{4 Feb, 2022}
\submitjournal{AJ}
\shorttitle{Atmospheric escape by flares}
\graphicspath{{./}{figures/}}
\usepackage{lineno}

\newcommand{\vplanet}{\texttt{VPLanet}\xspace}

\newcommand{\stellar}{\texttt{STELLAR}\xspace}
\newcommand{\atmesc}{\texttt{AtmEsc}\xspace}

\newcommand{\flare}{\texttt{FLARE}\xspace}

\begin{document}

\title{The Contribution of M-Dwarf Flares to the Thermal Escape of Potentially Habitable Planet Atmospheres}

\correspondingauthor{Laura Neves Ribeiro do Amaral}
\email{laura.nevesdoamaral@gmail.com}

\author[0000-0002-8341-0376]{Laura N. R. do Amaral}
\affiliation{Instituto de Ciencias Nucleares,
              Universidad Nacional Autónoma de México, Cto. Exterior S/N, C.U., Coyoacán, 04510 Ciudad de México, CDMX}

\author[0000-0001-6487-5445]{Rory Barnes}
\affiliation{Department of Astronomy, University of Washington,Seattle, WA 98105, USA}
\affiliation{NASA Virtual Planetary Laboratory Lead Team, USA}

\author[0000-0002-2240-2452]{Antígona Segura}
\affiliation{Instituto de Ciencias Nucleares,
              Universidad Nacional Autónoma de México, Cto. Exterior S/N, C.U., Coyoacán, 04510 Ciudad de México, CDMX}
\affiliation{NASA Virtual Planetary Laboratory Lead Team, USA}

\author[0000-0002-0296-3826]{Rodrigo Luger}
\affiliation{Center for Computational Astrophysics, Flatiron Institute, 162 Fifth Ave, New York, NY 10010, USA}
\affiliation{NASA Virtual Planetary Laboratory Lead Team, USA}

\begin{abstract}

The habitability of planets around M dwarfs ($\lesssim 0.5 M_\odot$) can be affected by the XUV (X rays + extreme UV) emission of these stars, with flares occasionally increasing the XUV flux by more than 2 orders of magnitude above quiescent levels. This wavelength range can warm and ionize terrestrial planets' upper atmospheres, which expands the planetary radius and promotes atmospheric loss. In this work, we study the contribution of the XUV flux due to flares on the atmospheric escape of Earth-like planets orbiting M dwarfs through numerical simulations. We considered the first Gyr of planets with initial surface water abundances between 1 and 10 terrestrial oceans (TO), a small primordial hydrogen envelope ($\le$ $10^{-3}$~$M_{\oplus}$), and with host star masses between 0.2 and 0.6 $M_{\odot}$. In this parameter range, we find that flares can remove up to two TO more than nonflaring stars, which, in some cases, translates to a doubling of the total water loss. We also find that flaring can increase atmospheric oxygen partial pressures by hundreds of bars in some cases. These results were obtained by adding a new module for flares to the \vplanet software package and upgrading its atmospheric escape module to account for Roche lobe overflow and radiation/recombination-limited escape.  

\end{abstract}


\section{Introduction} \label{sec:intro}

\quad Given the habitability of the Earth, it is reasonable to assume that potentially habitable planets may similarly consist of an iron-silicate interior and a liquid water surface ocean. For the purposes of this work, we will call such planets ``Earth-like'' planets.

To maintain surface water, the planet must have an atmosphere with greenhouse gases, which provides the pressure and temperature profile to maintain liquid water and avoid catastrophic water escape
\citep{1981Icar...48..150W,Kasting88,Barnes13,wordsworth2013water,luger2015extreme}.

As water is a solvent for a large number of biochemical reactions that facilitate prebiotic chemistry \citep{cockell2016habitability}, any process that can remove it is relevant for the search for life in the universe. Here we consider how XUV (X rays + extreme UV, 0.1--100 nm; \citealt{ribas2005evolution}) emission from stellar flares can photolyze atmospheric water and drive hydrogen escape to assess stellar flares impact on planetary habitability.

In order for a planet's surface temperature to be in the range for liquid water, the planet must receive an appropriate amount of energy from its star. The range of orbits around a particular star for which this condition is met for Earth-like planets is often called the habitable zone \citep[HZ;][]{Dole64,kasting1993habitable,kopparapu2013revised}. The M stars (0.07 -- 0.6 $M_\odot$) of the main sequence (also called M dwarfs) are currently the most observationally accessible targets for the search and characterization of Earth-like exoplanets \citep{billings2011astronomy,shields2016habitability, fujii2018exoplanet} because most stars are M dwarfs \citep{bochanski2010luminosity}, and Earth-like planets are relatively large and massive compared to their host star. Moreover, these stars stay on the  main sequence (MS) stage for 10$^{11}$ yr \citep{laughlin1997end,baraffe1998evolutionary,dotter2008dartmouth}, which is clearly much longer than the time it took for life to emerge on Earth. Preliminary reconnaissance of M dwarfs has revealed that 50$\%$ host an Earth-like planet  \citep{garrett2018planet,tuomi2019frequency}, so the Galaxy may be teeming with Earth-sized planets orbiting in the HZ of these low-mass stars.

\quad However, the habitability of M-dwarf planets can be severely compromised by certain characteristics of these stars. For example, M dwarfs may require billions of years to reach the main sequence \citep{hayashi1966evolution, laughlin1997end,baraffe1998evolutionary}, during which time these stars can follow the Hayashi track for over 1 billion years, with luminosities that can be over 1000 times larger than their zero-age main sequence luminosities. This change causes the HZ to move inward until the stars' cores begin to burn hydrogen \added{\citep{ramirez2014habitable,luger2015extreme,tian2015water}}. Once on the MS, the typical HZ for an M dwarf is at $\leq$ 0.25 au \citep{kasting1993habitable}. \citet{luger2015extreme} studied the impact of the pre-MS evolution (PMS) of the XUV radiation on M-dwarf planets and found that planets in the MS HZ might have experienced millions to billions of years of desiccating conditions, potentially rendering them dry and uninhabitable today. Additionally they showed that the desiccation process (water photolysis followed by hydrogen escape) could produce thousands of bars of oxygen. On the other hand, \citet{luger2015habitable} demonstrated that water escape can be suppressed if 1$\%$ of the planet's initial mass is in the form of a hydrogen atmosphere. In such a ``habitable evaporated core'' scenario, the hydrogen envelope insulates the water from the XUV radiation. \citet{barnes2016habitability} applied this model to Proxima Centauri b and found that it could lose five times the water content of the modern Earth during the pre-main sequence.

Another feature of M dwarfs is their strong variations in X-ray and extreme ultraviolet wavelengths, e.g. through stellar flares. Compared to the Sun, M-dwarf flares are more frequent and energetic (relative to the bolometric luminosity). While for the Sun the most energetic flares reach $10^{32}$ ergs and occur about once per solar cycle \citep{youngblood2017muscles}, for some M dwarfs, flare events with this same energy (or more) happen every day \citep{audard2000extreme, hawley2014kepler}. \citet{tilley2019modeling} showed that UV radiation and proton fluxes from repeated flaring can deplete the ozone layer of an Earth-like planet by 94$\%$ over 10 yr, so the oxygen left over from photolysis may be elemental. \citet{estrela2020surface} analyzed the impact of the UV radiation from flares on the potentially habitable planets of TRAPPIST-1 (an M8V star) and found that organisms that are nonresistant to UV could survive only if their habitats are deeper than 8 m below an ocean surface, or if the planet has an ozone layer. These studies highlight the importance of considering flares when assessing a planet's habitability.

\quad The XUV radiation emitted by M dwarfs (by chromospheric sources and flare events) ionizes and heats the exosphere of planetary atmospheres, slightly displacing it from hydrostatic equilibrium \citep{murray2009atmospheric}. This process generates an expansion of the atmosphere, where the exobase increases its radius, facilitating the escape because the cross section of the atmosphere increases during periods of high levels of XUV radiation (i.e., when the flaring is more frequent or when the star is more active) than in quiet periods \citep{france2020high}. Previous works have analyzed some of the impacts of XUV radiation from flares in the atmospheric escape of planets. A study by \citet{Atri2021} showed that XUV from flares (with a constant rate through the time) can drive, over 1 Gyr, escape rates of 3.38 $\times$ 10$^{-4}$ $M_{\oplus}$, 3.35 $\times$ 10$^{-4}$ $M_{\oplus}$, 1.46 $\times$ 10$^{-4}$ $M_{\oplus}$ for planets around stars of M4-M10, M0-M4, and FGK types, respectively. Similarly, a study by \citet{france2020high} analyzed the impact of three flares from Barnard's star in the atmosphere of an (unmagnetized) Earth-like planet and showed that these events can drive off the equivalent of $\sim 90$ Earth atmospheres in a period of 1 Gyr.

\quad Despite the impressive research into flaring and atmospheric escape, no study has yet evaluated how XUV emission from flares affects water loss and oxygen buildup in the atmospheres of planets orbiting M dwarfs. In this work, we present such analysis. To complete this task, we created a new module in the numerical code \vplanet \citep{barnes2020vplanet} that simulates the contribution of XUV by flares to the incident flux on an Earth-like planet and the resultant water photolysis, hydrogen escape, and oxygen accumulation. We employ a model for flares for stars between 0.2 and 0.6 $M_\sun$ \citep{davenport2019evolution} and have added this module, called \flare, to the {\vplanet codebase\footnote{\vplanet is publicly available at https://github.com/VirtualPlanetaryLaboratory/VPLanet.}.

\quad We divide this paper into the following sections. In section \ref{sec:model}, we describe and validate the \flare module. In section \ref{sec:simulation} we explain the simulated physical systems. Section \ref{sec:results} shows the results of the flare's influence when the model was applied to planets with a mass between 0.5 and 5 $M_{\oplus}$
around different M dwarf types and to four real planets. In section \ref{sec:discussion} we discuss the results, and in section \ref{sec:conclusion} we present the conclusions of our work. Note that the source code, simulation data, the input files, and the scripts that generate the figures and the figures are available online.
\footnote{
\url{https://github.com/lauraamaral/WaterEscapeFlares} and
\url{https://github.com/VirtualPlanetaryLaboratory/VPLanet}.}

\section{Model Description}
\label{sec:model}

In this section we present our physical models and numerical methods for simulating flares and atmospheric escape. To perform our simulations, we use the \vplanet software package that couples these processes. For this investigation, we have developed a new module for stellar flaring called \flare, so we also present the validation of this model in this section.

Figure \ref{fig:atmmodel} illustrates the physical processes in our model, which assumes all relevant planets form with an H envelope. The first panel, \ref{fig:atmmodel}.a, shows the system at the beginning of the simulation. Here, the XUV radiation (purple wiggly arrows) is interacting with the upper atmosphere of the planet (light blue). The model assumes the XUV radius $R_\mathrm{XUV}$, i.e. where the optical depth of the atmosphere to XUV photons is unity, is the same as the planet radius, including the atmosphere (see \citet{Salz2016} for discussion on the validity of this assumption). This choice means that XUV radiation does not penetrate deep into the atmosphere but follows the planetary radius boundary. This XUV energy interacting with the particles in the atmosphere increases its temperature and promotes its expansion, pushing out the exobase and increasing escape. Thus, as the radius of the planet decreases, the XUV radius approaches the planetary surface (brown), as seen in Figure \ref{fig:atmmodel}.b.

\begin{figure*}[htp!]
    \centering
    a {\includegraphics[width=0.18\textwidth]{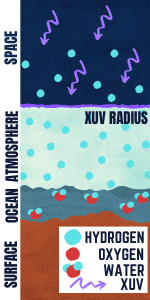}}
    b {\includegraphics[width=0.18\textwidth]{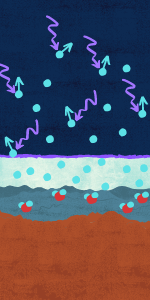}}
    c {\includegraphics[width=0.18\textwidth]{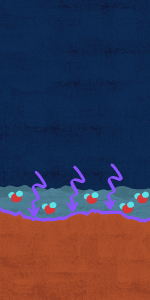}}
    d {\includegraphics[width=0.18\textwidth]{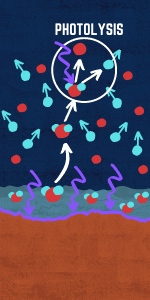}}
    e {\includegraphics[width=0.18\textwidth]{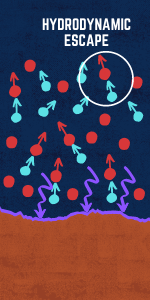}}
    \caption{Schematic of atmospheric and surface water escape in our model. The darkest blue represents space, and the lighter blue backgrounds, from top to bottom, represent the H envelope, the troposphere, and the ocean. The brown background (at the bottom of the figure) represents the planetary surface. The light blue dots are the hydrogen atoms, and the red dots the oxygen atoms. The purple arrows are the XUV radiation incoming from the planet. The horizontal purple curve is the XUV radius (where the optical depth of XUV is unity). See description in the text for more details. \href{https://github.com/lauraamaral/WaterEscapeFlares/tree/main/Plots/Scheme}{Plots/Scheme}. 
    }
    \label{fig:atmmodel}
\end{figure*}

\quad After the primordial atmosphere has escaped (Figure \ref{fig:atmmodel}.c), water in the stratosphere can absorb XUV photons and break apart into oxygen and hydrogen (see Fig.~\ref{fig:atmmodel}.d). This process can only occur if water can penetrate the tropopause, which we assume always occurs for planets orbiting interior to the HZ. If the hydrogen escape is vigorous enough, then oxygen can be dragged along with the hydrogen \citep{Hunten87}, as shown in Fig.~\ref{fig:atmmodel}.e. 

\subsection{\vplanet}

\quad To model the system, we use the software package \vplanet \citep{barnes2020vplanet}, which combines semi-analytical models to estimate the time evolution of parameters associated with planetary evolution and habitability. The code simulates how planetary system properties like stellar characteristics and orbital parameters affect the liquid water  on the planetary surface. It currently has 12 modules that calculate the contribution of the different planetary system properties. In the present work, we use three modules: \stellar and \atmesc, which simulate stellar evolution and atmospheric escape, respectively, as well as a new module, named \flare, to simulate stellar flaring. We next discuss each of these modules.

\vspace{-0.2cm}

\subsection{Atmospheric Escape: \atmesc}

To simulate atmospheric loss due to high-energy radiation, we use \vplanet's \atmesc module that accounts for the loss of a primordial hydrogen envelope, as well as water photolysis followed by hydrogen and oxygen escape.

\subsubsection{Envelope Loss}

\quad The initial version of \atmesc in \vplanet \citep{barnes2020vplanet} could only treat atmospheric escape as an energy-limited process, i.e. the flux of high-energy photons is the bottleneck for escape, not the supply of molecules and elements that are available to escape (diffusion-limited). Energy-limited escape generally takes the following form:

\begin{equation}
\dot{M}_{EL} = \frac{\epsilon_HF_\mathrm{XUV}R_\mathrm{XUV}}{4GM_\mathrm{XUV}K_\mathrm{tide}m_\mathrm{H}},
\label{eq:dfhdt}
\end{equation}
where $\dot{M}_{EL}$ is the energy-limited mass-loss rate, $F_\mathrm{XUV}$ is the XUV energy flux, $M_\mathrm{p}$ is the mass of the planet, $R_\mathrm{XUV}$ is the radius where XUV is absorbed and mass is escaping to space (we assume it equals the planet radius $R_\mathrm{p}$), and $\epsilon_\mathrm{XUV} \approx 0.1$ is the XUV absorption efficiency \citep{1981Icar...48..150W}. \cite{Erkaev07} introduced $K_\mathrm{tide}$ to approximate the decrease in escape velocity at the top of the planet's envelope due to the host star's gravity. They showed by scaling distances to the ratio of the Roche lobe radius $R_\mathrm{Roche}$ to the XUV radius $R_\mathrm{XUV}$:
\begin{equation}
    \chi = \frac{R_{Roche}}{R_{XUV}},
    \label{eq:chi}
\end{equation}
where
\begin{equation}
    R_{Roche} = \Big(\frac{M_p}{3M_*}\Big)^{1/3}a,
    \label{eq:roche}
\end{equation}
one could expand the gravitational potential to derive a relatively simple expression for the suppression of the local planetary gravity due to the nearby star. When $m_\mathrm{p} << M_\mathrm{*}$, as for the case of mini-Neptunes and main-sequence stars, \cite{Erkaev07} showed that
\begin{equation}
    K_{tide} \approx 1 - \frac{3}{2\chi} + \frac{1}{2\chi^3}.
    \label{eq:Ktide}
\end{equation}
For Earth-sized planets in the HZs of M dwarf stars, $K_\mathrm{tide}$ is typically between 0.9 and 0.99, so the effect is modest \citep[see also][]{luger2015habitable}, but its inclusion does increase the accuracy of the model.

For the radius of the planet, we interpolate the grids from the work of \cite{lopez2012thermal}, which are self-consistent models of ``mini-Neptunes'' over a range of masses and instellations. These grids take into account the evolution of the radius of the planet as the envelope contracts and entropy grows, independent of atmospheric escape. As mass is lost, \atmesc also interpolates between planetary masses. In all cases, we assume the water content does not affect planetary radius as the global ocean mass is always less than 0.2$\%$ the planetary mass for all cases we simulate.

Another study by \citet{turbet2020revised} showed that water-rich atmospheres can be an important factor to take into account in the mass-radius relation of water-rich rocky planets. In addition to the presence of water considered by \citet{turbet2020revised}, \citet{lopez2012thermal} also consider hydrogen and helium atmospheres. Thus, we decide to use the mass-radius relation from \citet{lopez2012thermal}.

We have now updated this module to include radiation/recombination-limited (RR) escape \citep{murray2009atmospheric} and Roche lobe overflow, which we will call ``Bondi-limited'' escape \citep{owen2016atmospheres}. The former occurs when the planetary radius is smaller than the Roche limit and the incident XUV flux in the planet is energetic enough to drive hydrogen ionization, which lowers the escape rate because some energy that could drive escape goes into ionization.

In SI units, RR-limited escape rate can be expressed as 
\begin{equation}
     \dot{M}_{RR} = 2.248 \times 10^6\Big(\frac{F_\mathrm{EUV}}{\textrm{W m$^{-2}$}}\Big)^{1/2}\Big(\frac{R_p}{R_\oplus}\Big)^{3/2}
     \textrm{kg s$^{-1}$},
     \label{eq:elim}
\end{equation}
where $\dot{M}_{RR}$ is the mass loss rate, and $F_\mathrm{EUV}$ is the extreme-UV energy flux incident on the planet. EUV observations are scarce \citep[e.g.][]{france19}   and to constrain the flux in this wavelength range, we must rely either on reconstructions using X-rays or FUV measurements \citep{sanzforcada2011,linsky2013}, stellar models \citep{Fontenla_2016,Peacock2020}, or use semiempirical methods \citep{Duvvuri_2021}. Given the uncertainties associated to the calculation of the EUV flux and the large differences among the fluxes obtained by the different approximations (see, for example, Fig. B1 in \citet{Peacock2020}, or section 7 in \citet{Duvvuri_2021}), we used the same flux for EUV and XUV. This means that our EUV flux is overestimated, which is consistent with our purpose of studying the worse case scenario for atmospheric loss. 

The RR limit occurs when the incident XUV reaches a critical value:
\begin{equation}
    F_{crit} = \Big(\frac{B}{A}\Big)^2,
    \label{eq:rrcrit}
\end{equation}
where
\begin{equation}
    A = \frac{\pi\epsilon_HR_{XUV}^3}{GM_pK_{tide}}
    \label{eq:rrA}
\end{equation}
and
\begin{equation}
    B = 2.248\times 10^6\Big(\frac{R_p}{R_\oplus}\Big)^{3/2} \textrm{kg}^\frac{1}{2} s^\frac{1}{2}.
    \label{eq:rrB}
\end{equation}
See \cite{luger2015habitable} for a derivation of these expressions. 

In extreme cases, such as after a planet-planet scattering event \citep{RasioFord96,LinIda97,Chatterjee08}, some planets may be so close to their star that the mass is stripped directly off the atmosphere by the stellar gravity. This process is typically called Roche lobe overflow. We model this process by assuming that mass is lost at the sound speed and call it ``Bondi-limited escape'', following \citet{owen2016atmospheres}. If we assume the atmosphere to be composed exclusively of molecular hydrogen that behaves as an ideal gas, then the sound speed at the top of the atmosphere can be expressed as
\begin{equation}
    c_s = \sqrt{\gamma k_bT_{eff}/m_H},
    \label{eq:soundspped}
\end{equation}
where $\gamma$ is 5/2 for an ideal gas, $k_\mathrm{b}$ is the Boltzmann constant, and $T_\mathrm{eff}$ is the effective temperature of the host star. Under these assumptions, we can recast Eq.~(4) of \cite{owen2016atmospheres} as

\begin{equation}
\begin{split}
    \dot{M}_{Bondi} \approx 1.9\times10^{15}\Big(\frac{M_p}{10M_\oplus}\Big)&\Big(\frac{T_{eff}}{5800 \textrm{K}}\Big)^{1/2} \\ &\Big(\frac{a}{0.1 \textrm{AU}}\Big)^{1/4}
    \Big(\frac{R_\odot}{R_*}\Big)^{1/4} 
    \textrm{kg s$^{-1}$} ,
    \label{eq:BondiLim}
\end{split}
\end{equation}
where we leave this expression as an approximation due to the assumptions we made, but in practice we set the Bondi-limited mass rate to be equal to the right-hand side of this expression.

For our calculations we employ all three of these models (when applicable) and allow the software to change escape regimes as the star and planet evolve based on instantaneous conditions. In Figures \ref{fig:RegimesLong} and \ref{fig:RegimesShort} we present an example of how the mass loss evolves assuming a solar mass star orbited by a $2M_\oplus$ planet that is half rock/half hydrogen by mass. The semi-major axis is 0.1 au, the eccentricity is 0, and $\epsilon_\mathrm{H} = 0.1$. Figure \ref{fig:RegimesLong} shows the long-term evolution, while Fig.~\ref{fig:RegimesShort} shows a zoom-in of the initial evolution of the envelope to highlight the Roche lobe overflow.

\begin{figure*}
\includegraphics[width=0.99\textwidth]{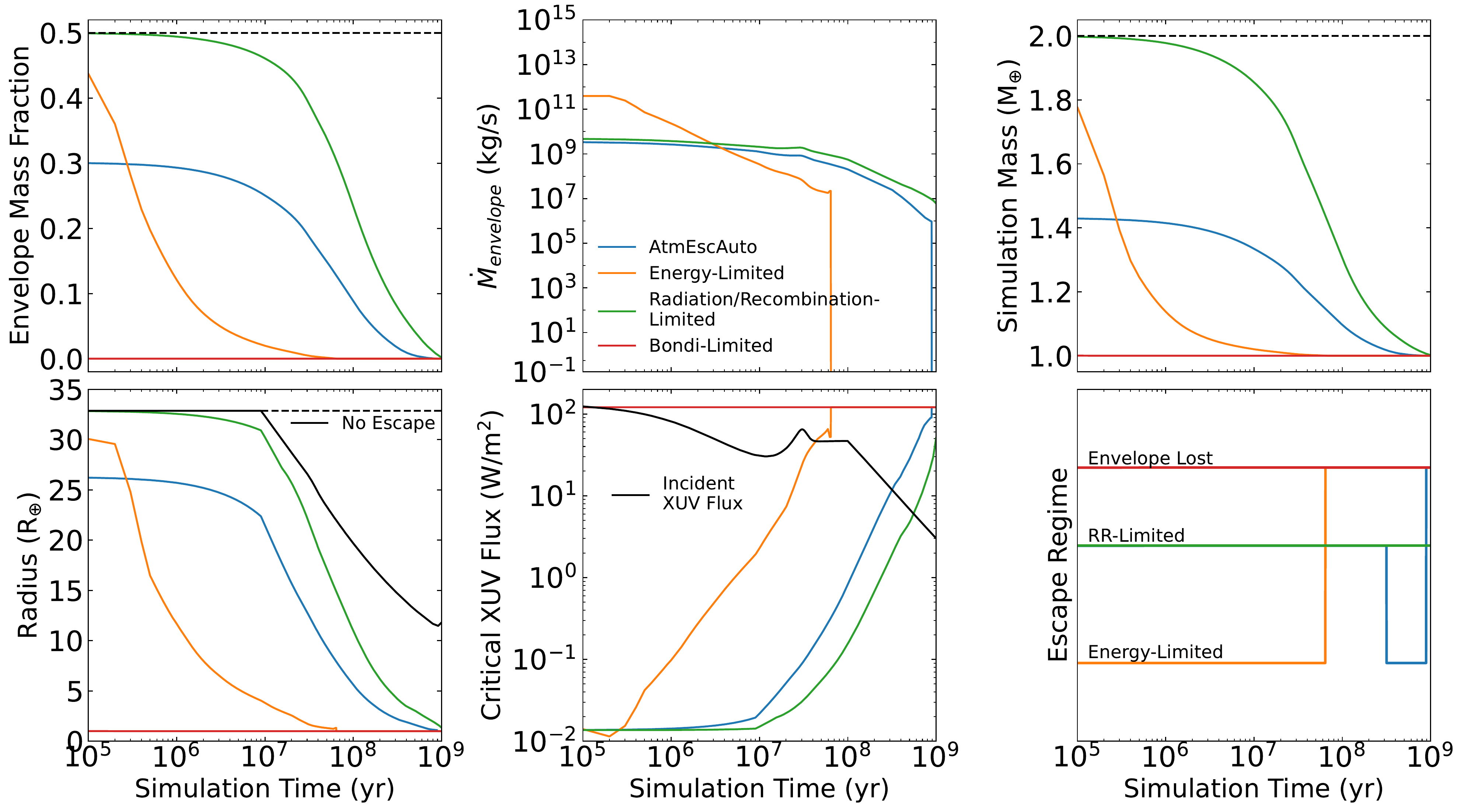}
\caption{Examples of the evolution of a planet experiencing different types of envelope
loss. In all panels, energy-limited
escape is represented by orange, radiation/recombination-limited by green, Bondi-limited by
red, and automatic ``AtmEscAuto'' by blue (the escape mechanism is determined by the instantaneous
planetary and environmental conditions). Top left: fraction of the planet's mass in the H envelope. Top center:
time derivative of the envelope mass. Top right: total planet mass. Bottom left:
planet radius. The ``No Escape'' curve (black) corresponds to a planet whose radius
contracts as the planet cools, not due to atmospheric escape. Bottom center:
critical XUV flux between energy-limited and radiation/recombination-limited escape.
Bottom right: the H envelope escape ``regime''. All planets start with the same initial conditions as described
in the main text. The blackdashed lines indicate the initial values
for the respective quantities. \href{https://github.com/VirtualPlanetaryLaboratory/vplanet/tree/main/examples/AtmEscRegimes}{VPLanet/examples/AtmEscRegimes}
}
\label{fig:RegimesLong}
\end{figure*}

\begin{figure*}
\includegraphics[width=0.99\textwidth]{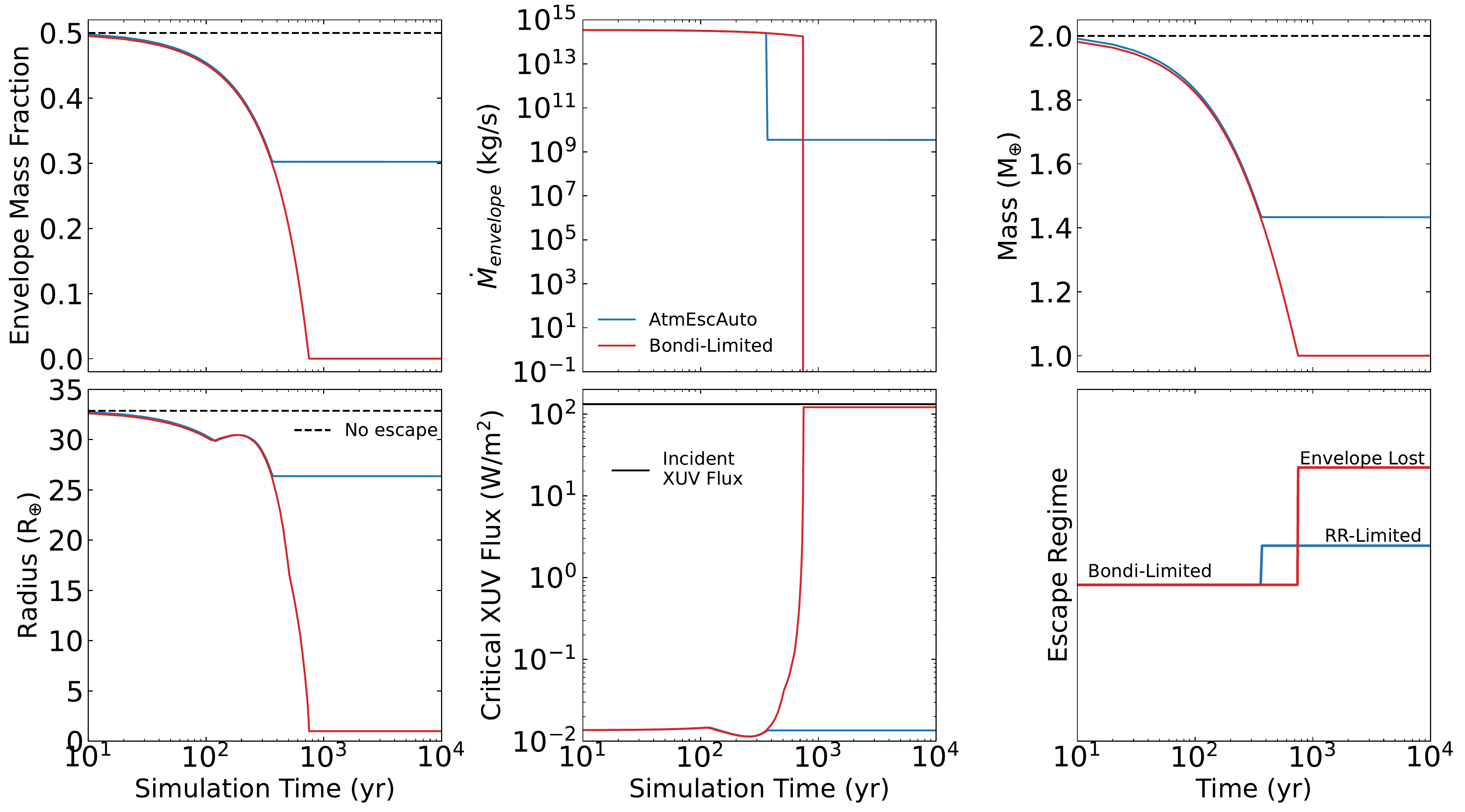}
\caption{First 10000 yr of atmospheric mass-loss for the Bondi-limited (red) and
automatic cases (blue) with the same layout as Fig.~\ref{fig:RegimesLong}.
Early on, the planetary
radius exceeds the Roche limit and both planets experience immense mass loss of order 0.01
Earth masses per year! After about 300 yr, the radius has contracted to be less than the Roche
limit, so the automatic model transitions from Bondi-limited escape to
radiation/recombination-limited escape (the XUV flux is too large for energy-limited escape.)
The Bondi-limited case, however, continues to lose massive amounts of H
from its envelope, completely stripping the envelope within 1000 yr.
\href{https://github.com/VirtualPlanetaryLaboratory/vplanet/tree/main/examples/AtmEscRegimes}{VPLanet/examples/AtmEscRegimes}
}
\label{fig:RegimesShort}
\end{figure*}

\subsubsection{Water and Oxygen Loss}

\quad After the hydrogen envelope has been removed, the XUV photons can begin to
dissociate the water molecules and drive more hydrogen escape and, in some cases, oxygen
escape. We use the \citet{Bolmont16} model for water escape as a function of $F_\mathrm{XUV}$ and refer the reader to \citet{barnes2020vplanet} for more details on how this model is implemented in \vplanet. We also include two ways in which water behaves when planets reach the HZ: they either stop losing water or they do not \citep[see][]{luger2015extreme, barnes2020vplanet, Becker20}.

The first case is a crude approximation of the setting of the conditions that allow the liquid water in the surface and avoid the catastrophic loss of water in the upper atmosphere \citep{wordsworth2013water}, and the second means planets are experiencing a runaway greenhouse. We use the ``optimistic'' HZ \citep{Kopparapu13} to determine when a planet is in the HZ.

\subsection{Quiescent Stellar Evolution: \stellar}

\quad The \stellar module \citep{barnes2020vplanet} simulates the time
evolution of stellar parameters such as the luminosity, radius, and effective temperature.
This module interpolates the model grids  of \citet{baraffe2015new} for stars between
0.08 and 1.3 $M_{\odot}$. However, the \citet{baraffe2015new} grids do not include XUV evolution,
which is poorly constrained for M-dwarf stars. Thus, we follow previous work and assume
the empirical relationship obtained for solar-type stars \citep{ribas2005evolution} applies to M dwarfs
as well \citep[see][]{luger2015extreme, Fleming20,Birky21}. In this model, the initial XUV
luminosity remains a constant fraction of the bolometric luminosity for a duration called
the ``saturation time,'' and afterwards the XUV fraction decays exponentially. See \cite{barnes2020vplanet} for more details.

A critical piece of our analysis is the time a planet spends interior to the HZ. If we assume the ``optimistic'' interior HZ limit from \cite{kopparapu2014habitable} denotes the inner edge of the HZ, then Fig.~\ref{fig:RGphase} shows the time a planet spends interior to the HZ as a function of host star mass, assuming the planet is on a static, circular orbit. Planets orbiting 0.2 $M_{\odot}$ stars remain in the runaway greenhouse phase almost 4 times longer than those orbiting 0.6 $M_{\odot}$ stars and are therefore expected to lose more water and possess more oxygen-rich atmospheres. The smaller variations in the boundary are due to structural and temporal variations related to the onset (or not) of convection in the stellar interiors  \citep{baraffe2018closer}. This result is consistent with \cite{luger2015extreme}.

\begin{figure}[!ht]
	   \centering
	     \includegraphics[width=0.75\textwidth]{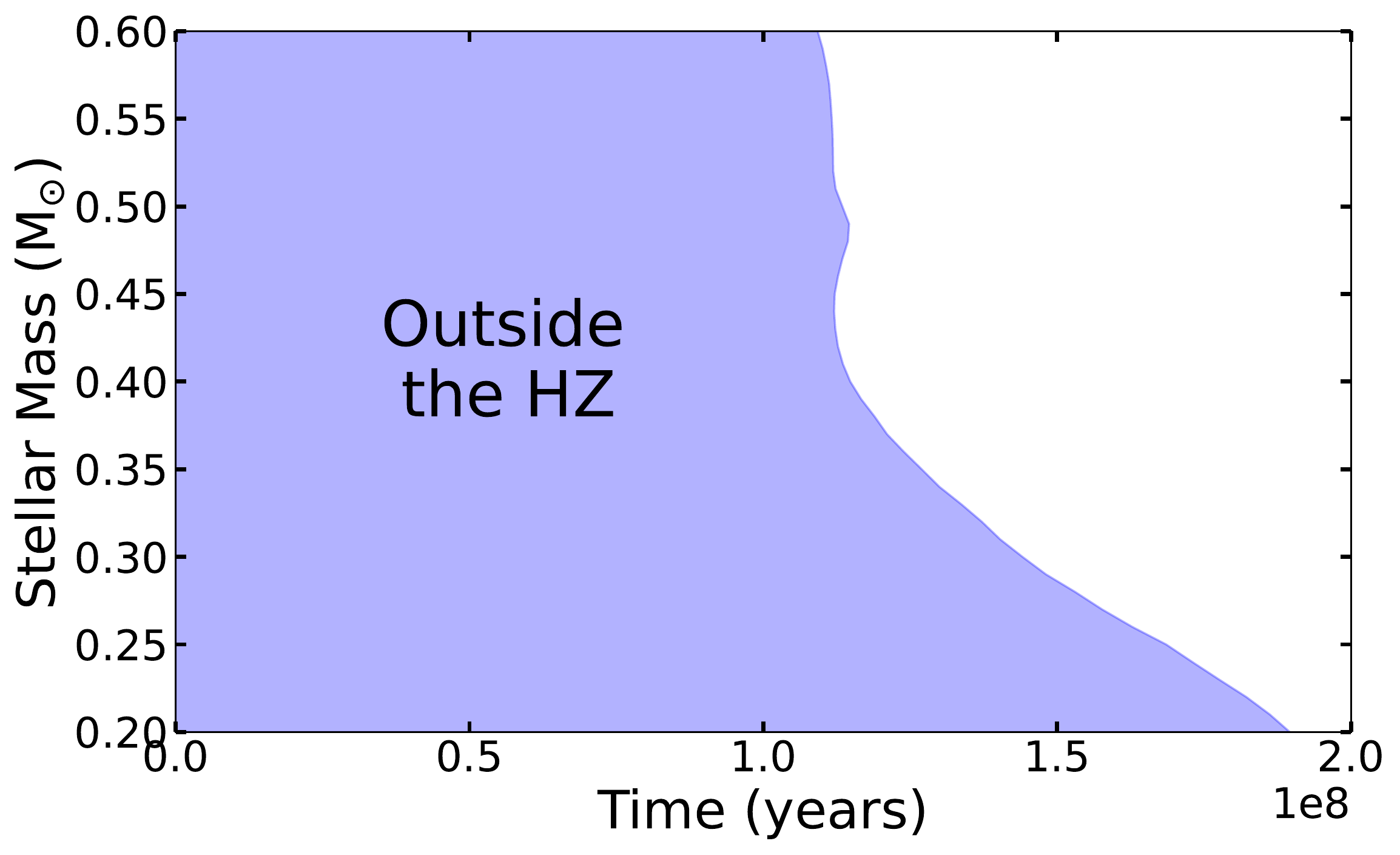}
\caption{Duration of the runaway greenhouse phase (blue shadow region) for the hypothetical planets as a function of stellar mass assuming stellar evolution from \cite{baraffe2015new} and the optimistic HZ limits from \cite{kopparapu2014habitable}.\href{https://github.com/lauraamaral/WaterEscapeFlares/tree/main/Plots/RGphase}{Plots/RGphase}.
}
\label{fig:RGphase}
\end{figure}

\subsection{XUV Flare Evolution: \flare }

Here we present our new \flare module for \vplanet. We first describe the mathematics of the model, followed by validation.

\subsubsection{Model Description}\label{sec:description}

While the \stellar module should capture most of the XUV luminosity of the star, it ignores
the contribution from flares. Therefore, we have updated \vplanet to include a new module
that estimates the time-averaged XUV luminosity due to stellar flares based on empirical data.
In general, the XUV luminosity contribution by flares ($L_{\mathrm{XUV},f}$) is given by

\begin{equation}
       L_{XUV,f} = \int_{E_{min}}^{E_{max}} \nu(E_{XUV,f})\ dE,
\label{eq:LXUVFlare}
\end{equation}
where $E_{\mathrm{XUV},f}$ is the XUV energy of the
flare, and $\nu$ is the flare rate per unit energy.

To solve Eq.~(\ref{eq:LXUVFlare}), we must know the flare rate, which depends on both the age
and stellar mass \citep[see, e.g.,][]{west2008constraining}. We use the canonical relation from
\citet{lacy1976uv}:

\begin{equation}
       \log_{10}(\nu) = \alpha\log_{10}(E_{Kepler})+\beta
\label{eq:ffd}
\end{equation}
with coefficients $\alpha$ and $\beta$ proposed by
\citet{davenport2019evolution} based on flare observations in the Kepler field.
The \citet{davenport2019evolution} model applies to active stars with masses between 0.2 and 1.88 $M_{\odot}$,
and stellar flares with energies (in the Kepler bandpass) between 10$^{33}$ and 10$^{38}$
ergs. We convert the Kepler bandpass to XUV using the conversion factors from
\citet[][Table 2]{osten2015connecting}.

\subsubsection{Model Validation}\label{sec:validation}

To validate our model, we present the flare rate evolution as a function
of the stellar age for a 0.5 $M_{\odot}$ star in Fig.~\ref{fig:daven}. This figure is indistinguishable from the top panel in Figure 10 from \cite{davenport2019evolution}. Thus, we conclude that we have successfully incorporated this flare model into \vplanet.

\begin{figure}[h!]
	   \centering
	     \includegraphics[width=0.6\textwidth]{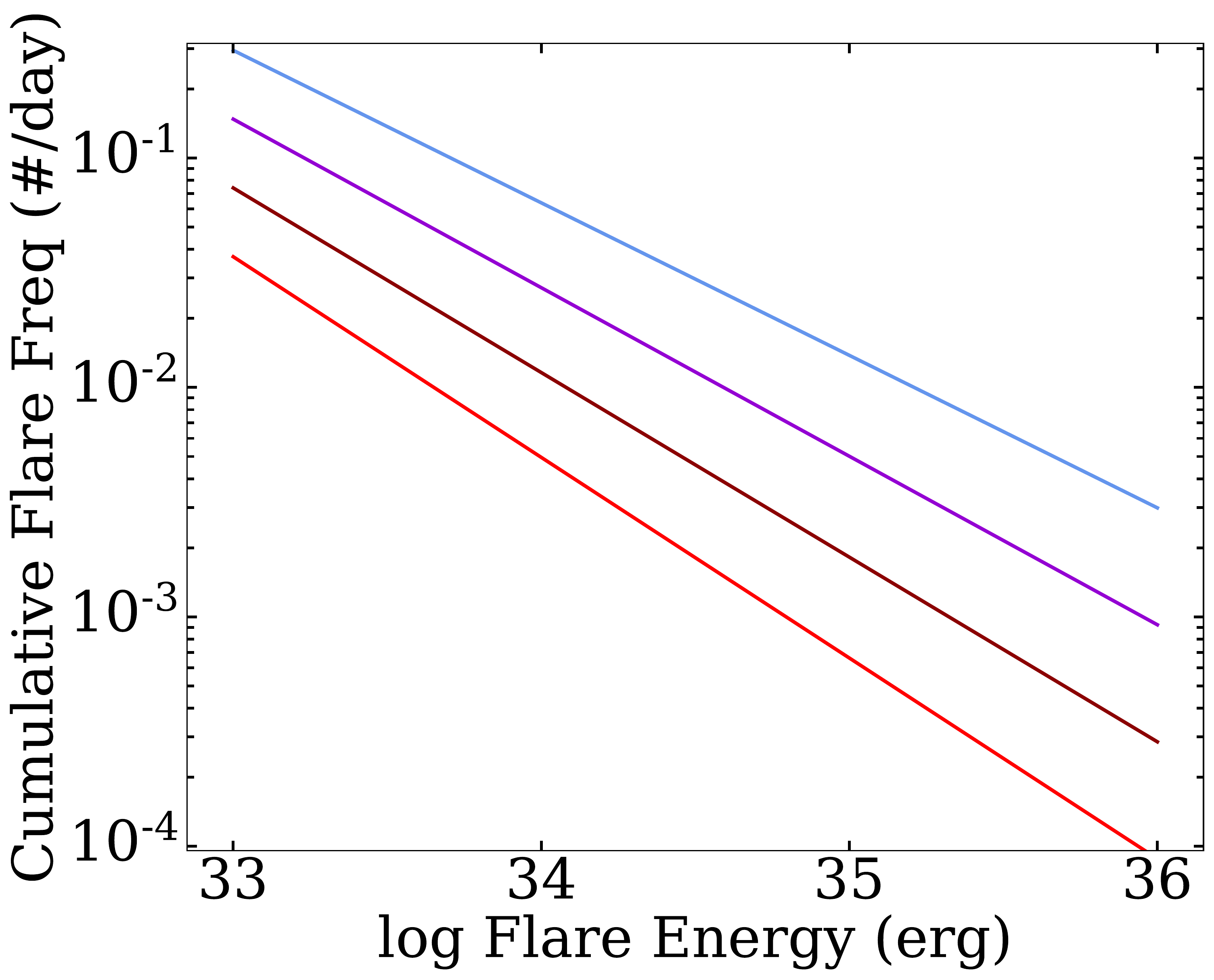}

\caption{Reproduction of Fig.~10 from \cite{davenport2019evolution} to validate the new \flare module.
The flare rate as a function of the flare energy for a 0.5 $M_{\odot}$ star is shown for ages of
1 Myr (blue), 10 Myr (purple), 100 Myr (brown), and 1 Gyr (red). \href{https://github.com/lauraamaral/WaterEscapeFlares/tree/main/Plots/MDwarfLuminosity/LumEvolFlareFFD}
{Plots/MDwarfLuminosity/LumEvolFlareFFD}.
}
\label{fig:daven}
\end{figure}

Before moving on, we provide a few characteristics of the \flare module. First, in Fig.~\ref{fig:ffd} we present two visualizations of how the flare frequency distribution (FFD) changes with time. The left panel shows the FFD for flares larger than the great AD Leo flare \citep{hawley1991great}, whose energy in the Kepler bandpass E$_\mathrm{kepler}$ would have been approximately $2.6 \times 10^{33}$ ergs.

This value was calculated using the information from \citet[Table 6A]{hawley1991great}, where we take the flux integrated over the time interval at the UV bandpass (326-394 nm) equal to 198 $\times$ 10$^{-8}$ ergs cm$^{-2}$, and that divided by the flux integrated over the time interval in the continuum equal to 1139 $\times$ 10$^{-8}$ ergs cm$^{-2}$. Doing a quickly calculation, we have that the energy in the UV bandpass E$_\mathrm{UV}$ divided by the energy in the continuum E$_\mathrm{continuum}$ is equal to 0.174. From \citet{hawley2014kepler}, we have that the ratio between the E$_\mathrm{Kepler}$ and E$_\mathrm{UV}$ is 0.65. Rearranging, we have that E$_\mathrm{Kepler}$ = 0.174/0.65 E$_\mathrm{continuum}$ = 0.26 E$_\mathrm{continuum}$. This value agrees with the same value that we can found in \citet{osten2015connecting}.

The \citet{davenport2019evolution} model predicts that planets orbiting early M dwarfs are more at risk of experiencing a flare of this magnitude than late M dwarfs. The right panel shows that the model predicts that less massive stars have a lower flare rate compared to more massive stars. This result seems to be inconsistent with other studies that find that flares from lower-mass stars are more powerful than those from larger stars \citep{hawley2014kepler} This discrepancy may be due to the small number of M dwarfs in the \cite{davenport2016kepler}  catalog ($<3\%$ were M dwarfs) that formed the basis for the \cite{davenport2019evolution} model. Despite this apparent inconsistency, the \cite{davenport2019evolution} model is derived from the largest sample of M dwarfs currently available, so we will use it here but explicitly note that future work may need to replace this model with a more robust approximation for M-dwarf flaring.

Next we consider how flaring and quiescent evolution combine to modify the average XUV luminosity of stars as a function of the mass and age. We simulated five M-dwarf stars with masses between 0.2 and 0.6 $M_{\odot}$ using the \stellar \citep{barnes2020vplanet} and \flare modules (this work). As shown in Figure \ref{fig:lum}, including flares increases the XUV luminosity but generally not by more than 10\% for any given star at a given time. We can see also that XUV luminosity by flares is more relevant to less massive stars after it enters the MS (see panel (e) and (f) of Figure \ref{fig:lum}). Also, panel (f) shows that the ratio between flare XUV and quiescent XUV increases with time. The quiescent XUV luminosity model we used comes from observational data of nonactive stars \citep{ribas2005evolution}, which show that quiescent XUV luminosity decreases with time. Hence as stars age, the XUV luminosity from flares becomes more important. The remaining panels provide additional depictions of the interplay of \stellar and \flare.

\begin{figure}
	   \centering
	     \includegraphics[width=0.99\textwidth]{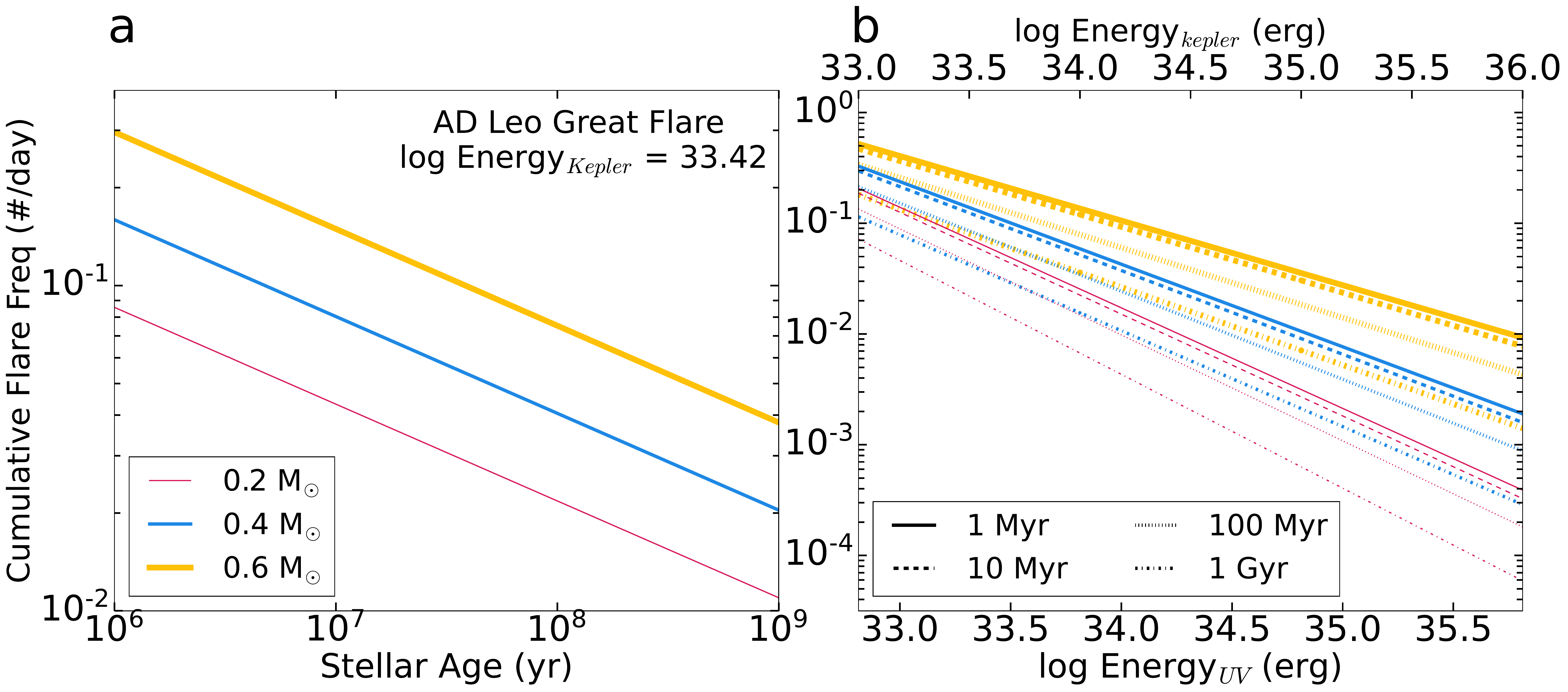}

\caption{Flare frequency distribution predicted by the \flare module. Panel (a) shows the FFD for flares with the energy of the AD Leo great flare (after converting the energy reported from \citet{hawley1991great} to the Kepler bandpass) as a function of stellar age for three different stellar masses. Panel (b) shows us the FFD for energies between 10$^{33}$ and 10$^{36}$ ergs for four different stellar ages. \href{https://github.com/lauraamaral/WaterEscapeFlares/tree/main/Plots/FfdMDwarfs}
{Plots/FfdMDwarfs}.
}
\label{fig:ffd}
\end{figure}

\begin{figure}
	   \centering
	     \includegraphics[width=0.9\textwidth]{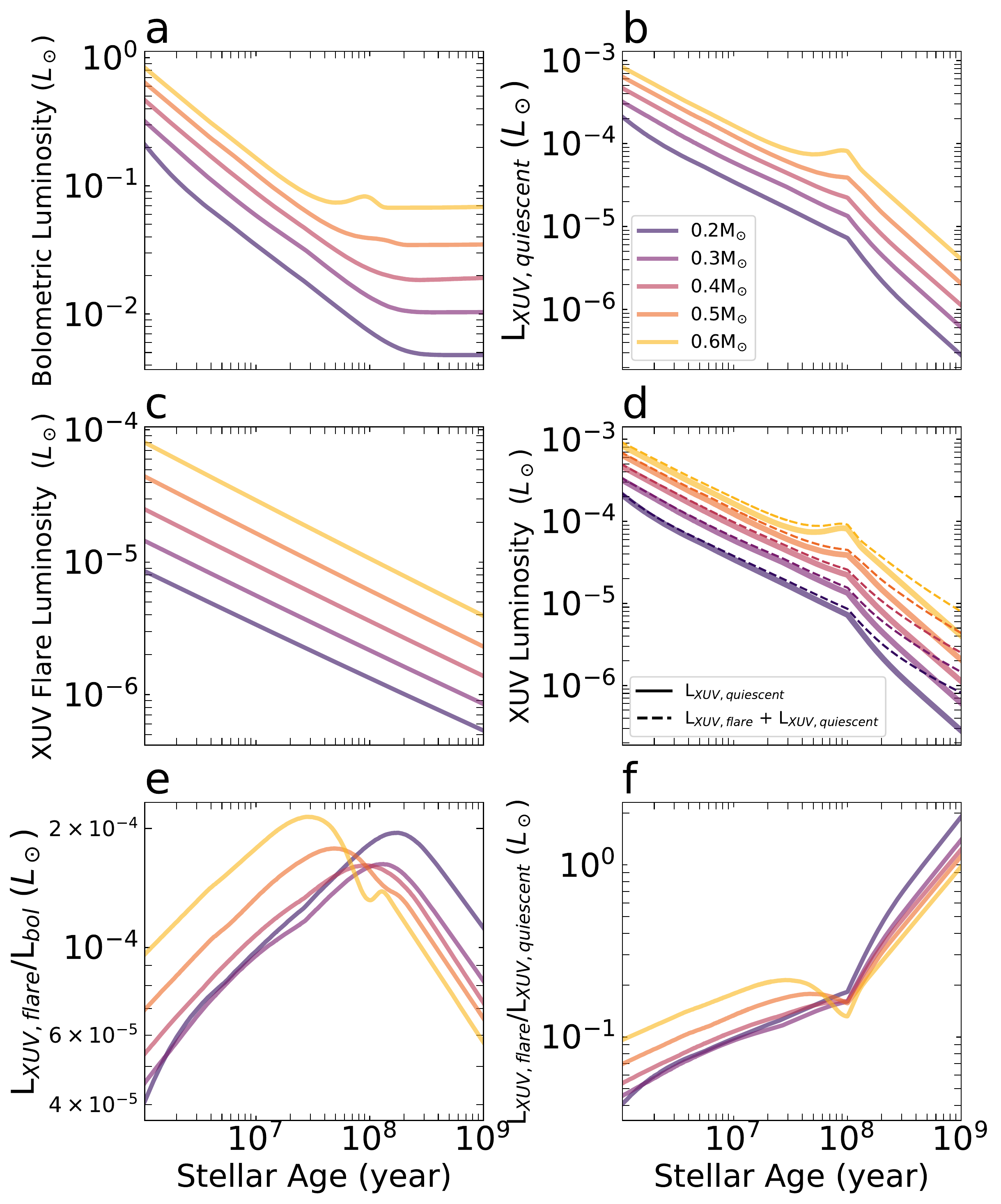}

\caption{Stellar evolution with and without flaring for stellar masses between 0.2 $M_{\odot}$ and 0.6 $M_{\odot}$. Panels (a) and (b) show the bolometric luminosity and XUV quiescent luminosity, respectively, while panels (c) and (d) show the XUV luminosity from flares and the total XUV luminosity with (dashed lines, i.e. the sum of parameters from panel (b) and (c)) and without (continuous lines) flares, respectively. Panel (e) shows the ratio between XUV luminosity from flares and bolometric luminosity, and panel (f) shows the ratio between XUV luminosity from flares and XUV quiescent luminosity. The colors represents different stellar mass as labeled in the legend. \href{https://github.com/lauraamaral/WaterEscapeFlares/tree/main/Plots/MDwarfLuminosity/LumEvolStellar}
{Plots/MDwarfLuminosity/LumEvolStellar}.
}
\label{fig:lum}
\end{figure}

\subsection{Simulations}
\label{sec:simulation}

To estimate the influence of XUV from flares on the time evolution of the atmosphere and water content of Earth-like planets, we simulated hypothetical and known planets. We assume the same atmospheric mass, thermosphere temperature, and absorption efficiency of XUV by hydrogen for all cases (see Table \ref{tab:general} and \ref{tab:real}). We consider flares with energies between 10$^{33}$ and 10$^{36}$ ergs, star masses between 0.2 and 0.6 $M_{\odot}$, surface initial water content between 1 and 10 terrestrial oceans (TO), planets with a mass between 0.5 and 5 $M_{\oplus}$, and hydrogen envelope masses of 0.001 $M_{\oplus}$. For each parameter combination, we perform one simulation for which water loss is halted in the HZ and one for which it is not.

The hypothetical set represents a parameter sweep that explores the general trends that can occur over a plausible range of initial surface water content, planet mass, and host-star mass.  The planet-star distance was selected to ensure that the planet would be interior to the HZ during the PMS of its host-star and enter the HZ when the star is 1 Gyr old by combining the runaway greenhouse limit \citep[][Eq. 5]{kopparapu2014habitable} with the predicted stellar properties at 1 Gyr from \stellar}. Furthermore, this synthetic group consisted of Earth-sized planets with the density of solids chosen from the \citet{sotin2007mass} mass-radius relationship. The initial conditions for these simulations are presented in Table \ref{tab:general}. In total, we simulated 25,160 scenarios. 

In the second group, we simulated the known exoplanets Kepler-1229 b, K2-72 e, TOI-700 d, and Kepler-1649 c, which have properties that are within the ranges of the hypothetical planets. We varied the water content and envelope mass but held fixed the values of the planetary and host-star masses, planetary radius, atmosphere temperature, and orbital parameters at their best-fit values. For the known planets, we only permit water loss prior to the planets reaching the HZ. For these planets we considered two distances: the reported semi-major axis for each planet (see Table \ref{tab:real}) and one calculated with the same procedure as for the hypothetical planets. These two sets of simulations enable direct comparisons between the results with the hypothetical planets. The parameters for these simulations are summarized in Table \ref{tab:real}, with values taken from the NASA Exoplanet Archive.

\begin{table}[htp!]
\caption{Parameters for the Hypothetical Cases.\label{tab:general}}
\centering
\setlength{\extrarowheight}{1pt}%
\begin{tabular}{lc}
\hline
\multicolumn{1}{c}{Parameter} & Value \\
\hline
\hline

Planet mass ($M_{\oplus}$)&  [0.5-5, 0.5] \\
Planet density (g cm$^{-3}$) & 4.85 - 7.54    \\
Envelope mass ($M_{\oplus}$) & 1.0E-3  \\
Surface water mass (TO) & [1-10,0.25] \\
XUV water escape efficiency & \citet{Bolmont16} \\
XUV hydrogen escape efficiency & 0.15 \\
Thermosphere temperature (K) & 400 \\
Semi-axis major$^{a}$ (AU) & [0.07306 - 0.283186]\\
Stellar mass $M_{\odot}$)  & [0.2-0.6, 0.025] \\
Saturated XUV luminosity fraction & 1.0E-3\\
XUV saturation time (Myr) &  100\\
Initial stellar age (Myr) & 1 \\
Flare energy (ergs) &  1.0E33 - 1.0E36 \\
Simulation time (Myr) & 1.0E3 \\
Time step$^{b}$ (yr) &  $\sim$ 0.4 - $\sim$ 1.6E3 \\
\vplanet modules & \scriptsize{\atmesc, \stellar, \flare}\\
\hline
\end{tabular}
\begin{flushleft}
\footnotesize{$^a$ Calculated using 1.053 $\times$ the distance of the inner limit of the HZ for runaway greenhouse from \citet[][Eq. 5]{kopparapu2014habitable}.\\
$^{b}$ Dynamically selected during the simulation.}
\end{flushleft}
\end{table}

\begin{sidewaystable}
\caption{Parameters used to simulate the final water content in the simulations of known planets.
\label{tab:real}}
\centering
\setlength{\extrarowheight}{1pt}%
\begin{threeparttable}
\begin{tabular}{lcccc}
\hline
\multicolumn{1}{c}{Parameter} & Kepler-1229 b & K2-72 e & TOI-700 d & Kepler-1649 c\\
\hline
\hline

 Planet mass$^{a}$ ($M_{\oplus}$)&  2.93 & 2.55 & 1.63984 & 1.2389   \\
 Planet radius ($R_{\oplus}$)& 1.34 & 1.29 & 1.144 & 1.06  \\
 Planet density (g cm$^{-3}$) & 6.69 & 6.52 & 6.018 & 5.715     \\
 Envelope mass ($M_{\oplus}$) & 1.0E-3  & 1.0E-3 & 1.0E-3 & 1.0E-3 \\
 Surface water (TO) & 1,10 & 1,10 & 1,10 & 1,10 \\
 XUV water escape efficiency & \citep{Bolmont16} & \citep{Bolmont16} & \citep{Bolmont16} & \citep{Bolmont16} \\
 XUV hydrogen escape efficiency & 0.15 & 0.15 & 0.15 & 0.15 \\
 Thermosphere temperature (K) & 400 & 400 & 400 & 400 \\
 Actual semi-major axis (AU) & 0.3006 & 0.106 & 0.1633 & 0.0827\\
 Modified semi-major axis$^{b}$ (AU) & 0.19685 & 0.102468 & 0.16247 & 0.07643\\
 Eccentricity  &  0.11 & 0 & 0.111 & 0\\
 Stellar mass ($M_{\odot}$)  & 0.480 & 0.271365 & 0.415 & 0.1977\\
 Saturated XUV luminosity fraction & 1.0E-3 & 1.0E-3 & 1.0E-3 & 1.0E-3 \\
 XUV saturation time (Myr) &  100&100&100&100\\
 Initial stellar age (Myr) & 1 & 1 & 1 & 1  \\
 Flare energy (ergs) &  11.0E33 - 1.0E36 &  1.0E33 - 1.0E36 &  1.0E33 - 1.0E36 &  1.0E33 - 1.0E36\\
 Simulation time (Myr) & 1.0E3 & 1.0E3 & 1.0E3 & 1.0E3 \\
 Time step$^{c}$ (yr) &  $\sim$4.8E-2 - 1.0E4 &  $\sim$6E-2 - 1.0E4 &   $\sim$5E-2 - 1.0E4 &  $\sim$6E-2 - 1.0E4 \\
\vplanet modules & \scriptsize \atmesc,  \stellar, \flare & \scriptsize \atmesc, \stellar, \flare & \scriptsize \atmesc, \stellar, \flare& \scriptsize \atmesc, \stellar, \flare \\
 Source (planetary radius, stellar & \scriptsize{\citet{torres2017validation} }& \scriptsize{\citet{dressing2017characterizing}} & \scriptsize{\citet{rodriguez2020first}} & \scriptsize{\citet{vanderburg2020habitable}} \\
  mass and semi-major axis) & & & &\\
\hline
\end{tabular}
\begin{flushleft}
  \footnotesize{
  \hspace{1.5cm}$^{a}$ The masses are calculated with the \citet{sotin2007mass} model for terrestrial planets.\\
  \hspace{1.5cm}$^{b}$ Calculated using \textbf{\replaced{1.052631579}{1.053}} $\times$ the distance of the inner limit of the HZ for runaway greenhouse from \citet[][Equation (5)]{kopparapu2014habitable}.\\
  \hspace{1.5cm}$^{c}$ Dynamically selected during the simulation.}
\end{flushleft}
\end{threeparttable}
\end{sidewaystable}

\section{Results}
\label{sec:results}

\subsection{Hypothetical Planets}

\begin{figure*}
  \includegraphics[width=0.98\textwidth]{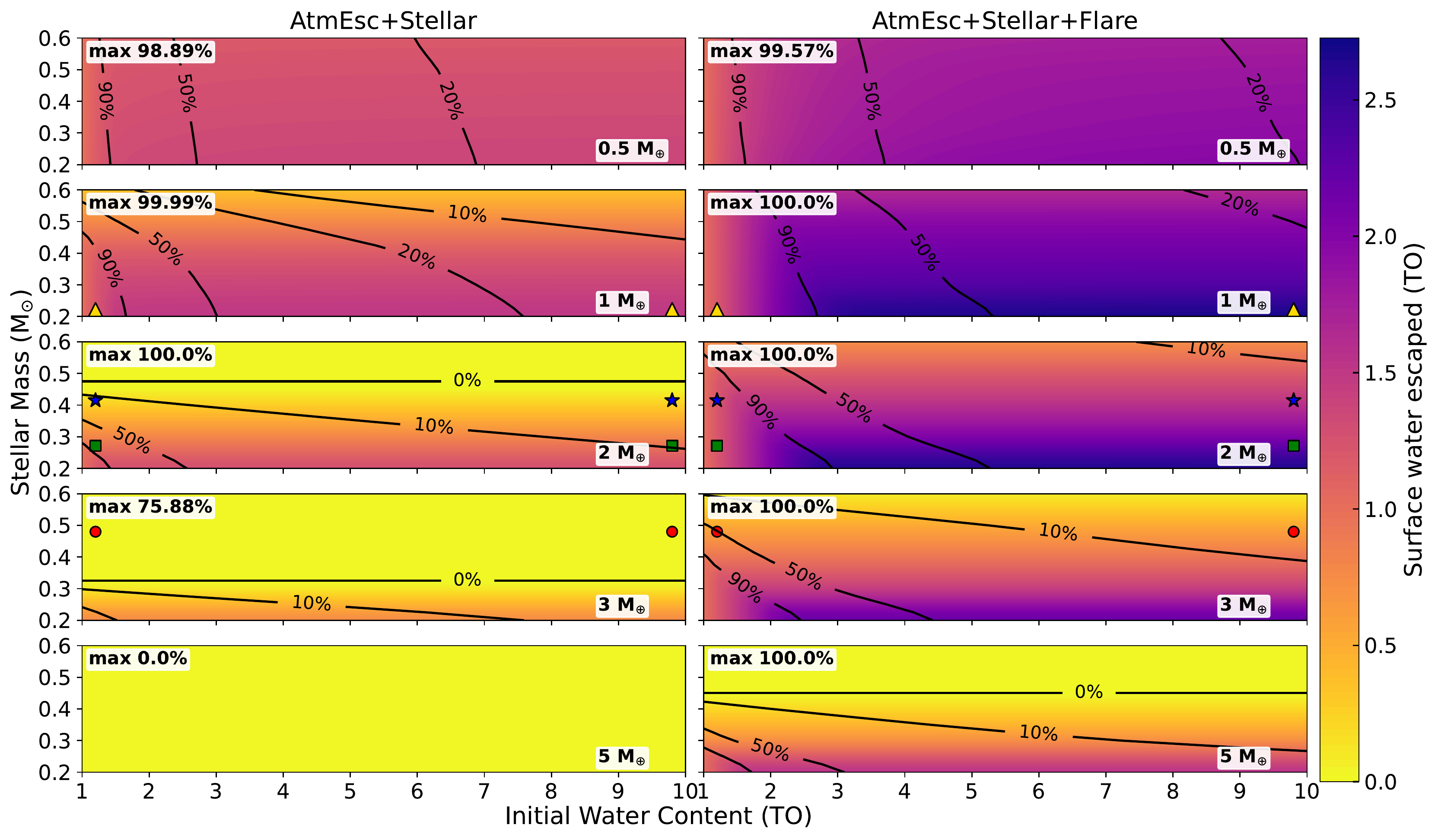}
  \includegraphics[width=0.98\textwidth]{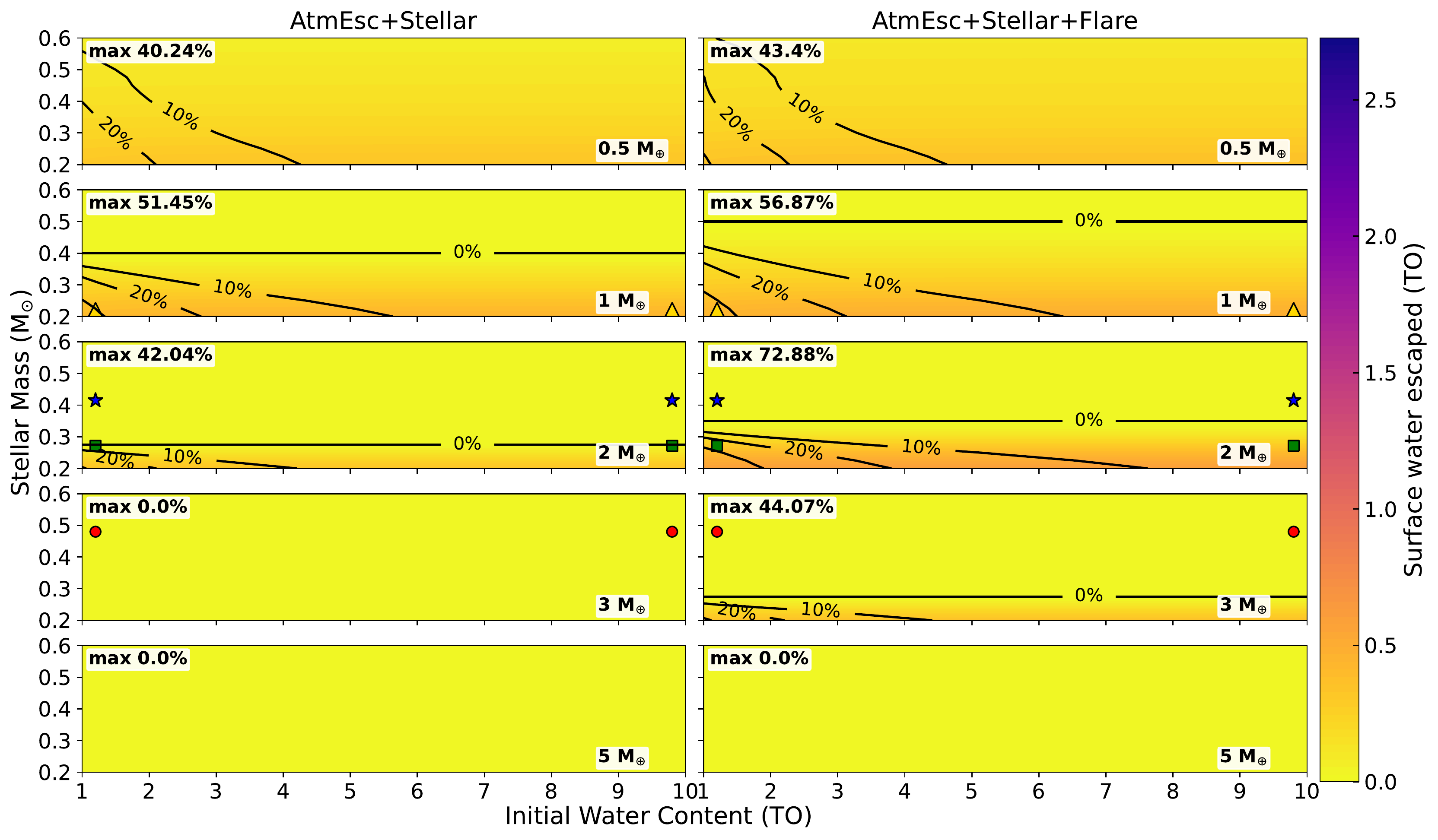}
  \caption{Percentage (contour lines) and absolute amount (shading) of water lost just by
  quiescent stellar XUV ({\tt\string AtmEsc} + {\tt\string Stellar} modules; left panels)
  and with the addition of
  flares ({\tt\string AtmEsc} + {\tt\string Stellar} + {\tt\string Flare} modules; right
  panels). In the upper panels, the planet continues to lose water even when in the HZ, while in the lower panels, water photolyzation halts in the HZ. The circle, square, star, and triangle represent
  Kepler-1229 b, K2-72 e, TOI-700 d, and Kepler-1649 c, respectively; see Table $\ref{tab:real}$.
  \href{https://github.com/lauraamaral/WaterEscapeFlares/tree/main/Plots/SurfaceWaterEscape/WaterEscapeAbsolute}
  {Plots/SurfaceWaterEscape/WaterEscapeAbsolute}.}
  \label{fig:LossInHZ}
\end{figure*}

Figure \ref{fig:LossInHZ} shows the amount of water lost from the hypothetical planets for four assumptions: with/without flares (right/left columns) and  with/without water loss in the HZ (lower/upper panels). The black lines represent the percentage of water lost with respect to the initial water content, and the symbols represent the approximate locations of the known planets discussed in Section \ref{subsec:known}. The amount of water lost is inversely proportional to the stellar and planetary mass in all four cases, but when flares are included, the amount of water lost increases by $\sim$ 0.6 TO when water is lost only in the HZ and by two TO when it is lost without restriction.

We find that water is only lost from massive planets when flares are included. This outcome is largely due to the more massive planets' ability to hold on to the hydrogen via their larger gravity. Without flares, the XUV flux is too low to drive significant mass loss, but flares provide enough XUV to drive additional loss, at least for the initial primordial envelope properties we assume here. We also find that the amount of water that escapes is independent of the initial water abundance.

\begin{sidewaysfigure}
  \includegraphics[scale=0.3]{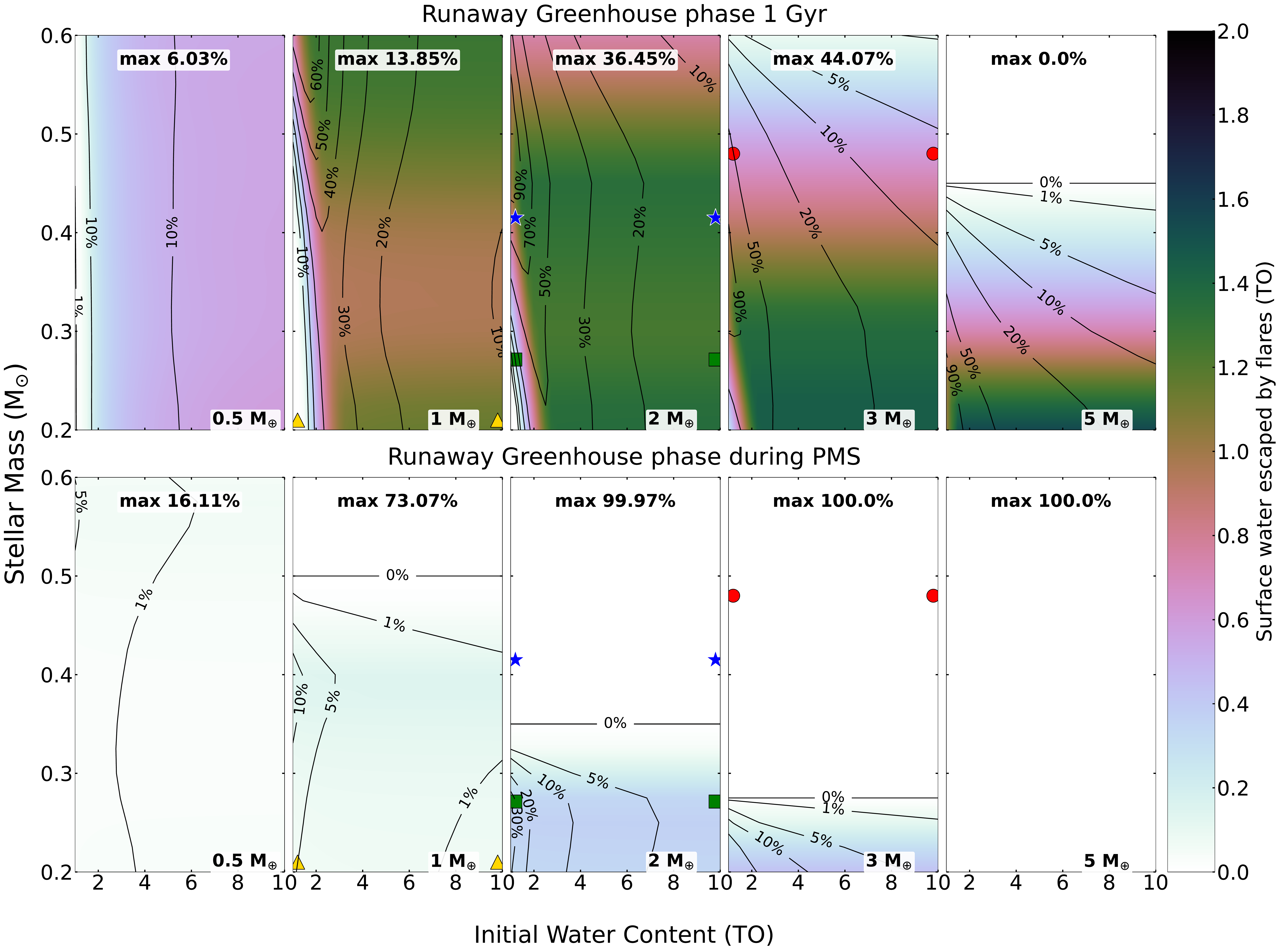}
  \caption{Flare only contribution to the water lost in percentage (contour lines) and absolute amount (shades) in TO, i.e., right panels minus left panels from \textbf{\replaced{Figures ref{fig:LossInHZ} and ref{fig:DoNotLossInHZ}}{Figure \ref{fig:LossInHZ}}}. In the top panels, the runaway greenhouse phase occurs throughout all the simulation. In the bottom panels, the runaway greenhouse effect stops when the planet enters the (optimistic) habitable zone. \href{https://github.com/lauraamaral/WaterEscapeFlares/tree/main/Plots/SurfaceWaterEscape/WaterEscapeTot}{Plots/SurfaceWaterEscape}.
  } 
  \label{fig:water_stop}
\end{sidewaysfigure}

To quantify the role of flares, we subtract the water loss from simulations that include flares from those that do not. Fig.~\ref{fig:water_stop} shows the results and includes both relative (contours) and absolute (colors) mass loss amounts. The top panels of Figure \ref{fig:water_stop} include water loss in the HZ, while the bottom panels do not. The first feature to note is that flares can remove up to two additional TO, a total desiccation of the surface water in some cases. When flares are not included, less water escapes from more massive planets orbiting more massive stars. On the other hand, cases with a shorter runaway greenhouse phase (bottom panels from Figure \ref{fig:water_stop}) or planets equal to or less massive than 2 $M_{\oplus}$ can only lose up to 36$\%$ more water due to flares. 

The biggest difference occurs for $\sim$3 $M_{\oplus}$ planets orbiting stars less massive than 0.3 $M_{\odot}$, as show in Figure \ref{fig:water_stop}. For the case with a runaway greenhouse in the HZ, the planets can lose up to an additional 0.6 TO of water than cases that not consider flares (a 44\% increase). For the cases where planets have a shorter period in the runaway greenhouse, i.e. only during PMS, the flares can remove up to two TO, or 100$\%$ of the surface water.

Note that for planets with small initial water content, the inclusion of flares does not affect water-loss rates because the quiescent XUV flux is sufficient for desiccation. In other words, these planets lose all their hydrogen and water regardless of stellar activity. In general, potentially habitable planets must form with at least four TO of water to be habitable after the PMS. With this value, the planets can keep (in a general way) approximately 50$\%$ of the amount of water they have at the beginning of their evolution.

\begin{sidewaysfigure}
  \includegraphics[scale=0.3]{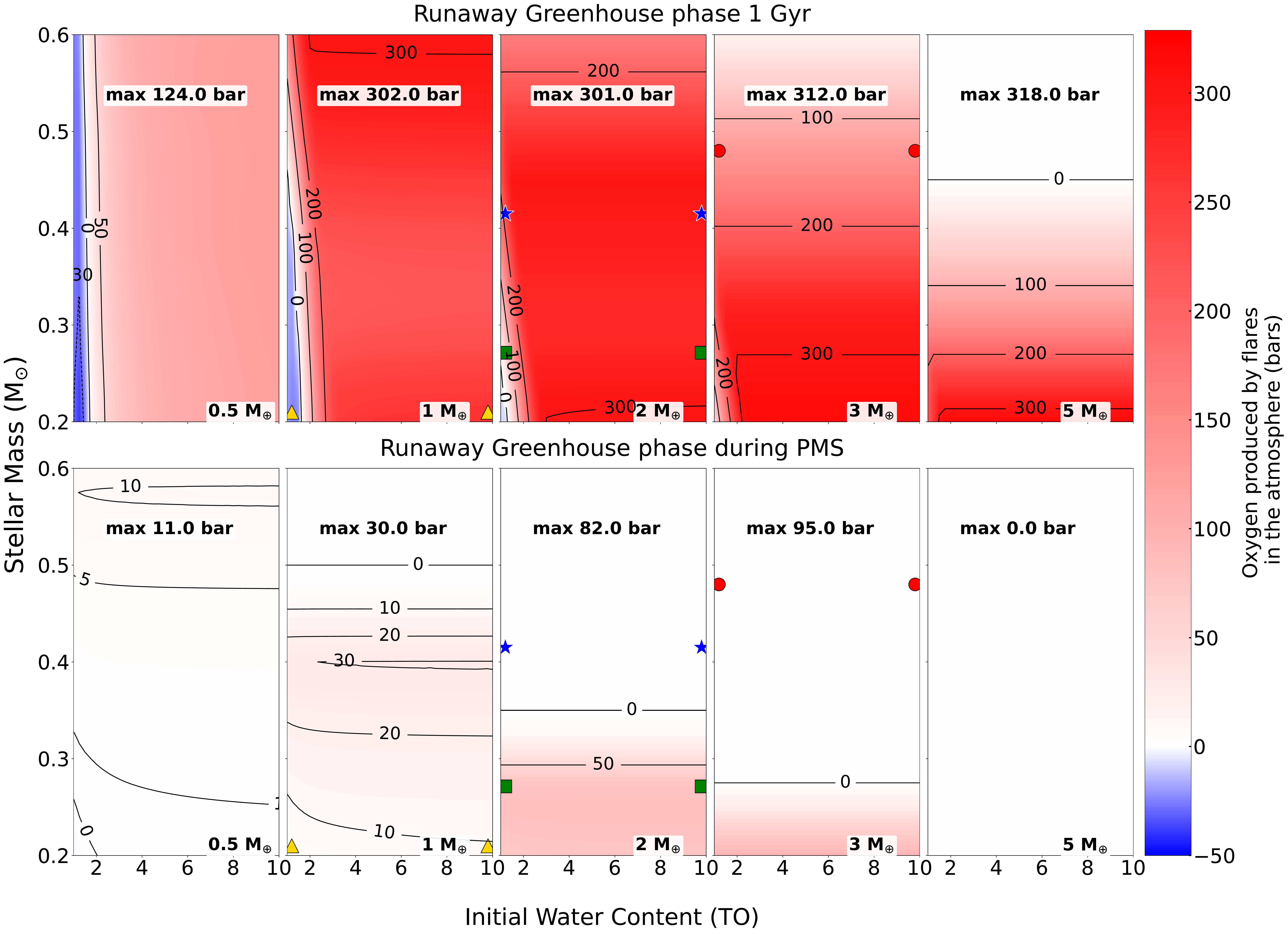}
  \caption{Amount of oxygen produced by flares (effect of \flare module) in the same format as Fig.~\ref{fig:water_stop}, but with different colors as shown in the color bar.  \href{https://github.com/lauraamaral/WaterEscapeFlares/tree/main/Plots/OxygenBuildUp}{Plots/OxygenBuildUp}.
  }
  \label{fig:oxygen_stop}
\end{sidewaysfigure}

Next we turn to the accumulation of the liberated oxygen in the atmosphere due to flares. In Fig.~\ref{fig:oxygen_stop} we plot the difference in final oxygen abundance between atmospheres that are exposed to flares versus those that are not in an analogous manner to Fig.~\ref{fig:water_stop}. The white regions of this figure are the cases where the oxygen produced by flares is equal to the quiescent case; red regions (positive values) show additional oxygen accumulation from flares; blue regions (negative values) show reduced oxygen abundance. 


Flares can generate up to 95 and 318 bars of additional oxygen when water photolysis is halted in the HZ and when it is not, respectively.
The blue regions do not represent cases in which less oxygen was produced, but rather where the XUV flux drives a more energetic flow of hydrogen that is able to drag along more oxygen. This effect can be seen by noting that the blue region overlaps with the fully desiccated regions of Fig.~\ref{fig:water_stop}, revealing that similar amounts of oxygen are produced but more oxygen escapes, resulting in less oxygen in the atmosphere at the conclusion of the simulations.

Even though we did not consider an oxygen sink in this work, we note that in a real Earth-like planet, the atmospheric oxygen can be removed in different ways, like metamorphism, weathering, and volcanism \citep{catling2017atmospheric}. As seen recently in previously works \citep{2016ApJ...829...63S,wordsworth2018redox, barth2021magma}, a magma ocean can also absorb oxygen efficiently. Nonetheless, all these studies still suggest that significant oxygen can accumulate in the atmosphere.

\vspace{-4cm}

\subsection{Known Planets}
\label{subsec:known}

\begin{figure}[H]
	   \centering
	     \includegraphics[width=0.99\textwidth]{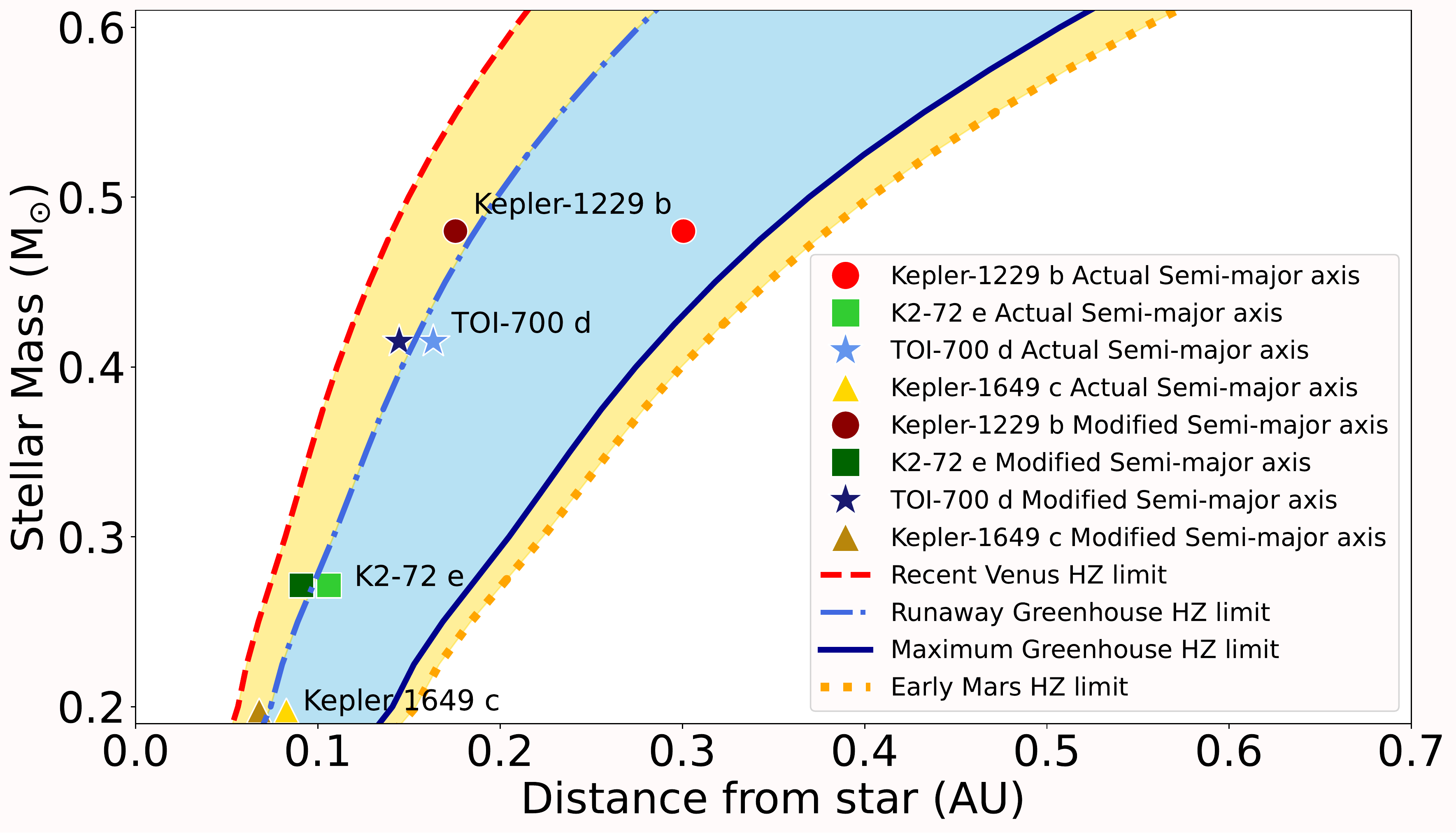}
\caption{Position of known planets with respect to the host star. The blue shading represents the conservative HZ, and the region between the red (dashed) and orange (dotted) lines are the optimistic HZ, as defined by \citet{kopparapu2014habitable}. The known planets are represented as circles (Kepler-1229 b), squares (K2-72 e), stars (TOI-700 d), and triangles (Kepler-1649 c). \href{https://github.com/lauraamaral/WaterEscapeFlares/tree/main/Plots/RealPlanets/RealPlanetsPosition}{Plots/RealPlanets/RealPlanetsPosition}.
}
\label{fig:HZ}
\end{figure}

Next we turn to the four real planets we selected in Section \ref{sec:validation}. For these simulations we assume that once a planet reaches the HZ, the water loss is halted.  Figure \ref{fig:HZ} shows the positions of these planets in the HZ. Figure \ref{fig:real} shows the time evolution of primordial envelope mass, surface water content, amount of atmospheric oxygen, optimistic HZ and semi-major axis, XUV-to-bolometric luminosity ratio, and incident XUV flux for each planet. In the top panels of both figures, this location is an extrapolation, once we did not simulate scenarios with an extended runaway greenhouse phase.

The simulations reveal that K2-72 e can lose its entire primordial H/He atmosphere when flares are included, regardless of the distance from the star.  However, this planet does not lose any water because it enters the HZ just before it loses all its primordial H/He atmosphere. About 600 Myr after the scenario with flares, the analog scenario with the quiescent XUV only also loses its primordial atmosphere. 

The Kepler-1649 c planet is the only one of these four simulated planets that loses water. Unlike K2-72 e, in Kepler-1649 c this process happens when the planet is placed closer to the star, with its known semi-major axis, regardless of flaring. The only difference is that, with flares, the primordial atmosphere escapes 30 Myr earlier.
 
Planet Kepler-1229 b does not lose its envelope even when flares are included, mostly due to its larger mass that helps to retain the hydrogen. TOI-700 d, even with less mass than K2-72 e, does not lose its primordial envelope either, and in this case, the larger orbital distance from the star further helps to prevent the atmospheric loss. All these results are in agreement with the hypothetical cases (see Fig. \ref{fig:LossInHZ} , \ref{fig:water_stop}, and, \ref{fig:oxygen_stop}), and confirm that the impact of flaring on envelope loss and desiccation depend strongly on the specific properties of a given planet.

\section{Discussion}
\label{sec:discussion}

\subsection{Dependence of the Stellar and Planetary Mass to the Final Water Reservoir}

After carrying out the simulations with the parameters shown in Table \ref{tab:general}, the values of the final amounts of water were taken for each simulated planet. With these data, the percentage of the water lost (compared to the initial value) was also calculated for each planet.

The hypothetical planets simulated in this work suggest strong correlations between stellar+planetary parameters and final water content. To quantify these relationships, we performed a Pearson's correlation test on those data. The Pearson correlation assumes linear dependency on pair of parameters. To use this test, we considered that the data are linearly correlated. The Pearson's coefficient varies from -1 (perfectly anticorrelated) to 1 (perfectly correlated). A Pearson coefficient equal to zero means that the data sets are uncorrelated. Figure \ref{fig:corr} shows the results of this analysis with respect to the percentage of surface water, for the four scenarios (water loss in the HZ and the inclusion of flares). Panels (b) and (d) include flares; panels (c) and (d) assume the planets do not lose water in the HZ (the runaway greenhouse phase occurs only during the PMS).

\begin{figure*}
	   \centering
	     \includegraphics[width=0.95\textwidth]{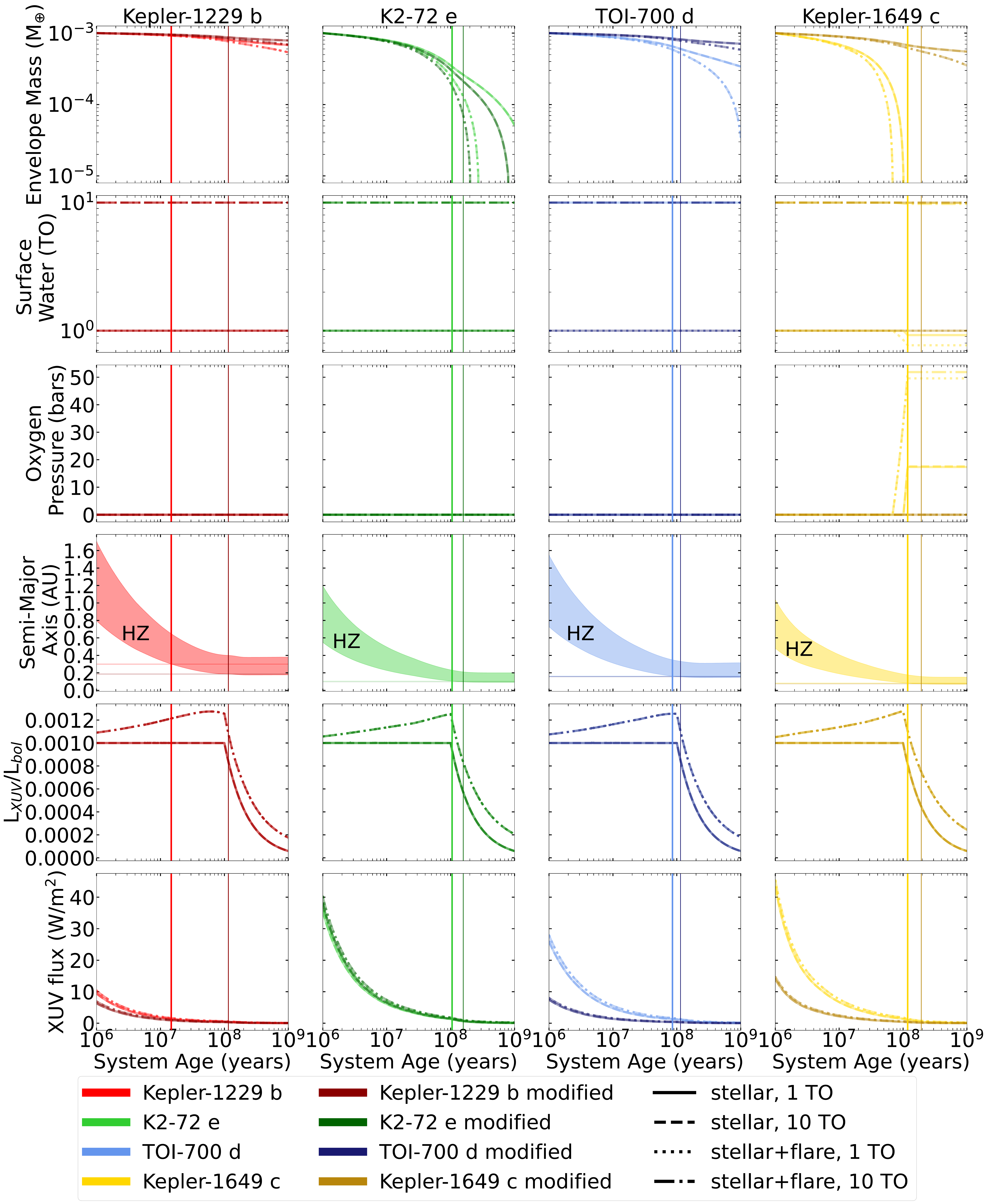}
\caption{Time evolution of planetary parameters for the known planets. From top to bottom: atmospheric mass, surface water, atmospheric oxygen, habitable zone position with respect to the host star, XUV and bolometric luminosity ratio, and XUV flux incoming the planet. \href{https://github.com/lauraamaral/WaterEscapeFlares/tree/main/Plots/RealPlanets/RealPlanetsSimulation}{Plots/RealPlanets/RealPlanetsSimulation}.
}
\label{fig:real}
\end{figure*}

In all cases, the Pearson test confirms that stellar and planetary masses are inversely correlated with the amount of water lost, i.e. less massive stars play a major role in  water loss. Although more massive stars have larger flare rates in our model, their PMS phase is shorter and thus less water is lost. For the scenario where we consider flares (panels (b) and (d) of Figure \ref{fig:corr}), the correlation between atmospheric loss with stellar mass increases compared to the cases without flares. When we assumed that the planet remains in a runaway greenhouse phase throughout 1 Gyr (panels (a) and (b) of Figure \ref{fig:corr}), the correlation between final water abundance and stellar mass decreases compared to planet mass because water loss is independent of the duration of the PMS. 

\begin{figure*}[htp!]
	\includegraphics[width=0.99\textwidth]{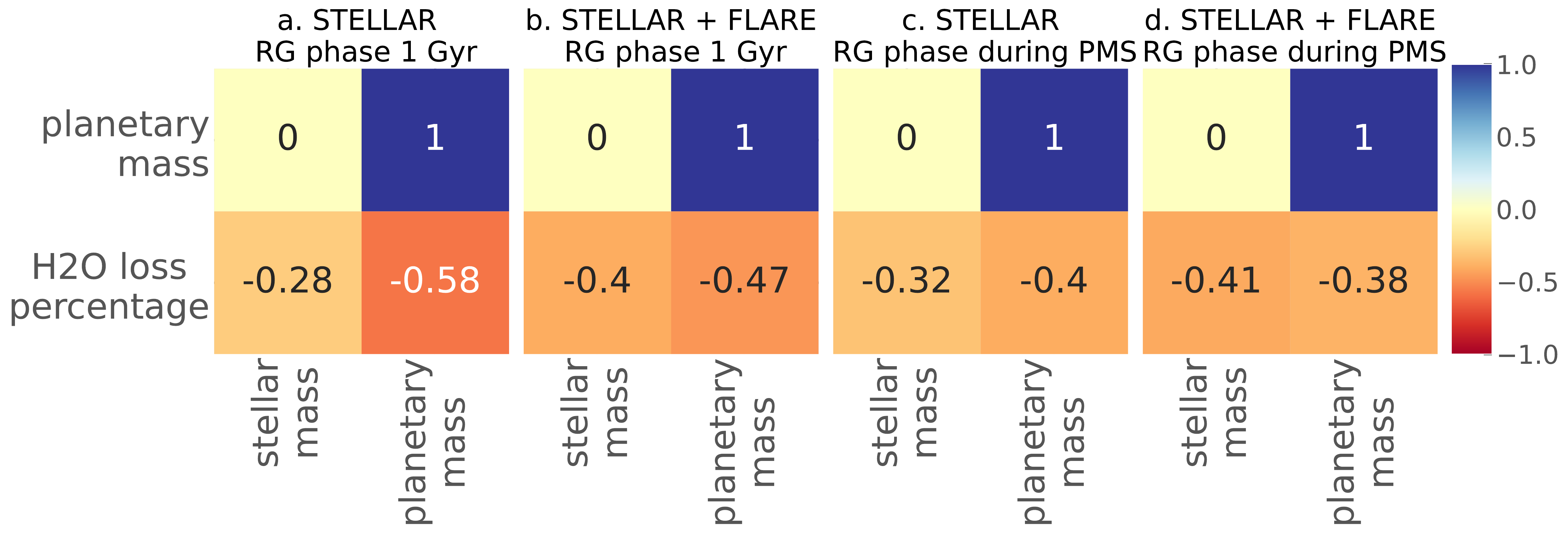}
\caption{Pearson correlation coefficient for final water mass for four different assumptions. Panel (a) assumes the planets lose water in the HZ and does not include flares. Panel (b) also assumes the planets lose water in the HZ but that the host star flares. Panels (c) and (d) assume water loss halts in the HZ, with panel (c) excluding flares and panel (d) including them. \href{https://github.com/lauraamaral/WaterEscapeFlares/tree/main/Plots/Correlation}{Plots/Correlation}.
}\label{fig:corr}

\end{figure*}

More massive planets possess stronger gravity and retain their atmosphere and water more easily than less massive planets, especially when the planet stays in the runaway greenhouse phase for 1 Gyr. If just the quiescent XUV of the star is considered, the correlation value is -0.58. When we add flares to the stellar evolution, the correlation drops to a value of -0.47.

This stellar mass trend depends on the length of the runaway greenhouse phase. When the runaway greenhouse phase is the same for all the stellar masses (panels (a) and (b)), the dependence on the stellar mass is weaker than when the  runaway greenhouse halts when the planet enters the HZ (panels (c) and (d)). Here the water lost depends on planetary mass almost equally (-0.4 and -0.38, respectively), but the stellar mass is more relevant when we consider flares (see panel (d)). In this case, the Pearson coefficient is -0.41 (panel (d)), compared to a value of -0.32 when we do not consider flares (panel (c)).

Finally, we note that the flare model from  \citet{davenport2019evolution} contains uncertainties that are not reflected in this analysis. In that work, the authors only considered a few M-dwarf stars in their sample, the observed flares have high energies ($>$31.5 ergs), and their analyzed stars are all active. These limitations likely make our results an upper limit of the simulated scenarios. Future flare observations of M dwarfs (e.g., achieved with the Transiting Exoplanet Survey) will likely require modifications of the FFD as a function of stellar mass and time. When such an effort has been completed, the results presented here should be revisited.

\vspace{-0.1cm}

\subsection{Impact of Flares on Known Planets}

Our simulations of known planets revealed a wide range of plausible evolutionary trajectories; however, the range is likely underestimated. We assumed that the planets orbit active stars, but that is probably not the case for many of them.
TOI-700 is a slow rotator \citep[period = 54 days;][]{newton2017halpha,gilbert2020first,rodriguez2020first}, indicating this star is not currently active. Kepler-1229 has a rotation period of 17.63 days \citep{torres2017validation}, so it could be active, but like K2-72 and Kepler-1649, it has no activity or flares reported in the literature. However, these observations are all for their current conditions and, since stellar activity decreases with time \citep{west2008constraining}, our simulations may still be representative of their early evolution. Otherwise, the results presented here can be considered an upper limit (worst case) of the environment that these planets are exposed to.

\citet{cohen2020space} found that TOI-700 d is not in an extreme environment compared to Earth, even considering a strong coronal mass ejection (CME) event. \citet{dong2020atmospheric} found that the planet's atmosphere could be stripped within 1 Gyr if the planet is unmagnetized. This result is consistent with our study, as shown in Figure \ref{fig:real}. If we extrapolate the results, TOI-700 d's envelope mass could be lost just after 1 Gyr. Note, however, that our simulations also predict that water can be retained, even when considering flares. Even with flares, the only case that experienced significant water lost was Kepler-1649 c, removing 0.16 TO more than quiescent evolution alone. This result is likely due to the short orbital distance and small stellar mass, which increase the time the planet is interior to the HZ.

\section{Conclusions}
\label{sec:conclusion}

We simulated the XUV emission of M-dwarf stars to estimate the atmospheric escape and oxygen buildup on synthetic and known Earth-size planets. For the first set of simulations, we  modeled a range of parameters, such as the stellar mass, planetary mass, and initial water abundance, to estimate the trajectories that permit water on the planetary surfaces today, i.e. a habitable planet. We find that many planets, including some that are known, could be habitable today. However, we also note that other processes can affect habitability, such as tidal forces, planetary magnetic field, CMEs, or proton events. We also did not consider radiative cooling mechanisms, such as from CO$_2$, in the simulated atmospheres. These are important features to the atmospheric escape when we are analyzing planetary habitability but are beyond the scope of this work and should be explored with future research. Because of all these points, our results should be considered preliminary, especially for the known planets.

As part of this work, we added the \citet{davenport2019evolution} flare frequency distribution model to the \vplanet software package as a module we call \flare. We validated this model by reproducing the results in \citet{davenport2019evolution}. This module is now part of this open-source project and available for community use.

We found that flares add about 10\% more XUV emission to M dwarfs over the quiescent stellar levels, which can remove up to an additional two TO of surface water compared to quiescent stars, at least for Earth-like planets. We assumed a primordial hydrogen envelope mass of 0.001 $M_{\oplus}$, so actual water amounts lost will change for different envelope assumptions. We also found that flares play their most significant role in water escape for planets between 2 and 5 $_M{\oplus}$. Furthermore, the final water content depends more strongly on the stellar mass than the planetary mass when water loss halts once the planet reaches the HZ. However, if the planets continue losing water inside the HZ, then planetary mass is more important. These trends hold for our simulations of known planets. Future space missions such as the James Webb Space Telescope and the ground-based Extremely Large
Telescope may be able to test the predictions presented here, possibly even leading to the discovery of an inhabited exoplanet.

\vspace{2cm}

R.B. acknowledges support from NASA grants 80NSSC20K0229 and the NASA Virtual Planetary Laboratory Team through grant number 80NSSC18K0829. L.N.R.A. and A.S. acknowledge the support of UNAM DGAPA PAPIIT project IN110420. L.N.R.A. thanks CONACYT’s graduate scholarship program for its support. We thank David Fleming for developing the automatic mass loss functionality and Caitlyn Wilhelm for developing the vspace and the multiplanet codes that facilitated our parameter sweeps.

\facility{Exoplanet Archive} 
\vspace{5mm}

\software{\vplanet \citep{barnes2020vplanet}
          }

\bibliographystyle{apalike}
\bibliography{main}

\begin{thebibliography}{}

\bibitem[{Atri} and {Mogan}, 2021]{Atri2021}
{Atri}, D. and {Mogan}, S. R.~C. (2021).
\newblock {Stellar flares versus luminosity: XUV-induced atmospheric escape and
  planetary habitability}.
\newblock {\em \mnras}, 500(1):L1--L5.

\bibitem[Audard et~al., 2000]{audard2000extreme}
Audard, M., G{\"u}del, M., Drake, J.~J., and Kashyap, V.~L. (2000).
\newblock Extreme-ultraviolet flare activity in late-type stars.
\newblock {\em The Astrophysical Journal}, 541(1):396.

\bibitem[Baraffe and Chabrier, 2018]{baraffe2018closer}
Baraffe, I. and Chabrier, G. (2018).
\newblock A closer look at the transition between fully convective and partly
  radiative low-mass stars.
\newblock {\em Astronomy \& Astrophysics}, 619:A177.

\bibitem[Baraffe et~al., 1998]{baraffe1998evolutionary}
Baraffe, I., Chabrier, G., Allard, F., and Hauschildt, P. (1998).
\newblock Evolutionary models for solar metallicity low-mass stars:
  mass-magnitude relationships and color-magnitude diagrams.
\newblock {\em Astronomy and Astrophysics}, 337:403--412.

\bibitem[Baraffe et~al., 2015]{baraffe2015new}
Baraffe, I., Homeier, D., Allard, F., and Chabrier, G. (2015).
\newblock New evolutionary models for pre-main sequence and main sequence
  low-mass stars down to the hydrogen-burning limit.
\newblock {\em Astronomy \& Astrophysics}, 577:A42.

\bibitem[Barnes et~al., 2016]{barnes2016habitability}
Barnes, R., Deitrick, R., Luger, R., Driscoll, P.~E., Quinn, T.~R., Fleming,
  D.~P., Guyer, B., McDonald, D.~V., Meadows, V.~S., Arney, G., et~al. (2016).
\newblock The habitability of proxima centauri b i: evolutionary scenarios.
\newblock {\em arXiv preprint arXiv:1608.06919}.

\bibitem[Barnes et~al., 2020]{barnes2020vplanet}
Barnes, R., Luger, R., Deitrick, R., Driscoll, P., Quinn, T.~R., Fleming,
  D.~P., Smotherman, H., McDonald, D.~V., Wilhelm, C., Garcia, R., et~al.
  (2020).
\newblock Vplanet: The virtual planet simulator.
\newblock {\em Publications of the Astronomical Society of the Pacific},
  132(1008):024502.

\bibitem[Barnes et~al., 2013]{Barnes13}
Barnes, R., Mullins, K., Goldblatt, C., Meadows, V., Kasting, J., and Heller,
  R. (2013).
\newblock Tidal venuses: Triggering a climate catastrophe via tidal heating.
\newblock {\em Astrobiology}, 13:225--250.

\bibitem[Barth et~al., 2021]{barth2021magma}
Barth, P., Carone, L., Barnes, R., Noack, L., Molli{\`e}re, P., and Henning, T.
  (2021).
\newblock Magma ocean evolution of the trappist-1 planets.
\newblock {\em Astrobiology}, 21(11):1325--1349.

\bibitem[{Becker} et~al., 2020]{Becker20}
{Becker}, J., {Gallo}, E., {Hodges-Kluck}, E., {Adams}, F.~C., and {Barnes}, R.
  (2020).
\newblock {A Coupled Analysis of Atmospheric Mass Loss and Tidal Evolution in
  XUV Irradiated Exoplanets: The TRAPPIST-1 Case Study}.
\newblock {\em \aj}, 159(6):275.

\bibitem[Billings, 2011]{billings2011astronomy}
Billings, L. (2011).
\newblock Astronomy: Exoplanets on the cheap.
\newblock {\em Nature News}, 470(7332):27--29.

\bibitem[{Birky} et~al., 2021]{Birky21}
{Birky}, J., {Barnes}, R., and {Fleming}, D.~P. (2021).
\newblock {Improved Constraints for the XUV Luminosity Evolution of
  Trappist-1}.
\newblock {\em Research Notes of the American Astronomical Society}, 5(5):122.

\bibitem[Bochanski et~al., 2010]{bochanski2010luminosity}
Bochanski, J.~J., Hawley, S.~L., Covey, K.~R., West, A.~A., Reid, I.~N.,
  Golimowski, D.~A., and Ivezi{\'c}, {\v{Z}}. (2010).
\newblock The luminosity and mass functions of low-mass stars in the galactic
  disk. ii. the field.
\newblock {\em The Astronomical Journal}, 139(6):2679.

\bibitem[{Bolmont} et~al., 2017]{Bolmont16}
{Bolmont}, E., {Selsis}, F., {Owen}, J.~E., {Ribas}, I., {Raymond}, S.~N.,
  {Leconte}, J., and {Gillon}, M. (2017).
\newblock {Water loss from terrestrial planets orbiting ultracool dwarfs:
  implications for the planets of TRAPPIST-1}.
\newblock {\em \mnras}, 464:3728--3741.

\bibitem[Catling and Kasting, 2017]{catling2017atmospheric}
Catling, D.~C. and Kasting, J.~F. (2017).
\newblock Atmospheric evolution on inhabited and lifeless worlds.
\newblock {\em Atmospheric Evolution on Inhabited and Lifeless Worlds}.

\bibitem[{Chatterjee} et~al., 2008]{Chatterjee08}
{Chatterjee}, S., {Ford}, E.~B., {Matsumura}, S., and {Rasio}, F.~A. (2008).
\newblock {Dynamical Outcomes of Planet-Planet Scattering}.
\newblock {\em \apj}, 686:580--602.

\bibitem[Cockell et~al., 2016]{cockell2016habitability}
Cockell, C.~S., Bush, T., Bryce, C., Direito, S., Fox-Powell, M., Harrison, J.,
  Lammer, H., Landenmark, H., Martin-Torres, J., Nicholson, N., et~al. (2016).
\newblock Habitability: a review.
\newblock {\em Astrobiology}, 16(1):89--117.

\bibitem[Cohen et~al., 2020]{cohen2020space}
Cohen, O., Garraffo, C., Moschou, S.-P., Drake, J.~J., Alvarado-G{\'o}mez, J.,
  Glocer, A., and Fraschetti, F. (2020).
\newblock The space environment and atmospheric joule heating of the habitable
  zone exoplanet toi 700 d.
\newblock {\em The Astrophysical Journal}, 897(1):101.

\bibitem[Davenport, 2016]{davenport2016kepler}
Davenport, J.~R. (2016).
\newblock The kepler catalog of stellar flares.
\newblock {\em The Astrophysical Journal}, 829(1):23.

\bibitem[Davenport et~al., 2019]{davenport2019evolution}
Davenport, J.~R., Covey, K.~R., Clarke, R.~W., Boeck, A.~C., Cornet, J., and
  Hawley, S.~L. (2019).
\newblock The evolution of flare activity with stellar age.
\newblock {\em The Astrophysical Journal}, 871(2):241.

\bibitem[{Dole}, 1964]{Dole64}
{Dole}, S.~H. (1964).
\newblock {\em {Habitable planets for man}}.
\newblock Rand Corporation.

\bibitem[Dong et~al., 2020]{dong2020atmospheric}
Dong, C., Jin, M., and Lingam, M. (2020).
\newblock Atmospheric escape from toi-700 d: Venus versus earth analogs.
\newblock {\em The Astrophysical Journal Letters}, 896(2):L24.

\bibitem[Dotter et~al., 2008]{dotter2008dartmouth}
Dotter, A., Chaboyer, B., Jevremovi{\'c}, D., Kostov, V., Baron, E., and
  Ferguson, J.~W. (2008).
\newblock The dartmouth stellar evolution database.
\newblock {\em The Astrophysical Journal Supplement Series}, 178(1):89.

\bibitem[Dressing et~al., 2017]{dressing2017characterizing}
Dressing, C.~D., Vanderburg, A., Schlieder, J.~E., Crossfield, I.~J., Knutson,
  H.~A., Newton, E.~R., Ciardi, D.~R., Fulton, B.~J., Gonzales, E.~J., Howard,
  A.~W., et~al. (2017).
\newblock Characterizing k2 candidate planetary systems orbiting low-mass
  stars. ii. planetary systems observed during campaigns 1--7.
\newblock {\em The Astronomical Journal}, 154(5):207.

\bibitem[Duvvuri et~al., 2021]{Duvvuri_2021}
Duvvuri, G.~M., Pineda, J.~S., Berta-Thompson, Z.~K., Brown, A., France, K.,
  Kowalski, A.~F., Redfield, S., Tilipman, D., Vieytes, M.~C., Wilson, D.~J.,
  Youngblood, A., Froning, C.~S., Linsky, J., Loyd, R. O.~P., Mauas, P.,
  Miguel, Y., Newton, E.~R., Rugheimer, S., and Schneider, P.~C. (2021).
\newblock Reconstructing the extreme ultraviolet emission of cool dwarfs using
  differential emission measure polynomials.
\newblock {\em The Astrophysical Journal}, 913(1):40.

\bibitem[{Erkaev} et~al., 2007]{Erkaev07}
{Erkaev}, N.~V., {Kulikov}, Y.~N., {Lammer}, H., {Selsis}, F., {Langmayr}, D.,
  {Jaritz}, G.~F., and {Biernat}, H.~K. (2007).
\newblock {Roche lobe effects on the atmospheric loss from ``Hot Jupiters''}.
\newblock {\em A\&A}, 472:329--334.

\bibitem[Estrela et~al., 2020]{estrela2020surface}
Estrela, R., Palit, S., and Valio, A. (2020).
\newblock Surface and oceanic habitability of trappist-1 planets under the
  impact of flares.
\newblock {\em Astrobiology}, 20(12):1465--1475.

\bibitem[{Fleming} et~al., 2020]{Fleming20}
{Fleming}, D.~P., {Barnes}, R., {Luger}, R., and {VanderPlas}, J.~T. (2020).
\newblock {On the XUV Luminosity Evolution of TRAPPIST-1}.
\newblock {\em \apj}, 891(2):155.

\bibitem[Fontenla et~al., 2016]{Fontenla_2016}
Fontenla, J.~M., Linsky, J.~L., Witbrod, J., France, K., Buccino, A., Mauas,
  P., Vieytes, M., and Walkowicz, L.~M. (2016).
\newblock {SEMI}-{EMPIRICAL} {MODELING} {OF} {THE} {PHOTOSPHERE},
  {CHROMOPSHERE}, {TRANSITION} {REGION}, {AND} {CORONA} {OF} {THE} m-{DWARF}
  {HOST} {STAR} {GJ} 832.
\newblock {\em The Astrophysical Journal}, 830(2):154.

\bibitem[France et~al., 2020]{france2020high}
France, K., Duvvuri, G., Egan, H., Koskinen, T., Wilson, D.~J., Youngblood, A.,
  Froning, C.~S., Brown, A., Alvarado-G{\'o}mez, J.~D., Berta-Thompson, Z.~K.,
  et~al. (2020).
\newblock The high-energy radiation environment around a 10 gyr m dwarf:
  Habitable at last?
\newblock {\em The Astronomical Journal}, 160(5):237.

\bibitem[France et~al., 2019]{france19}
France, K., Fleming, B.~T., Drake, J.~J., Mason, J.~P., Youngblood, A.,
  Bourrier, V., Fossati, L., Froning, C.~S., Koskinen, T., Kruczek, N., Lipscy,
  S., McEntaffer, R., Romaine, S., Siegmund, O. H.~W., and Wilkinson, E.
  (2019).
\newblock {The extreme-ultraviolet stellar characterization for atmospheric
  physics and evolution (ESCAPE) mission concept}.
\newblock In Siegmund, O.~H., editor, {\em UV, X-Ray, and Gamma-Ray Space
  Instrumentation for Astronomy XXI}, volume 11118, pages 38 -- 51.
  International Society for Optics and Photonics, Society of Photographic
  Instrumentation Engineers.

\bibitem[Fujii et~al., 2018]{fujii2018exoplanet}
Fujii, Y., Angerhausen, D., Deitrick, R., Domagal-Goldman, S., Grenfell, J.~L.,
  Hori, Y., Kane, S.~R., Pall{\'e}, E., Rauer, H., Siegler, N., et~al. (2018).
\newblock Exoplanet biosignatures: observational prospects.
\newblock {\em Astrobiology}, 18(6):739--778.

\bibitem[Garrett et~al., 2018]{garrett2018planet}
Garrett, D., Savransky, D., and Belikov, R. (2018).
\newblock Planet occurrence rate density models including stellar effective
  temperature.
\newblock {\em Publications of the Astronomical Society of the Pacific},
  130(993):114403.

\bibitem[Gilbert et~al., 2020]{gilbert2020first}
Gilbert, E.~A., Barclay, T., Schlieder, J.~E., Quintana, E.~V., Hord, B.~J.,
  Kostov, V.~B., Lopez, E.~D., Rowe, J.~F., Hoffman, K., Walkowicz, L.~M.,
  et~al. (2020).
\newblock The first habitable-zone earth-sized planet from tess. i. validation
  of the toi-700 system.
\newblock {\em The Astronomical Journal}, 160(3):116.

\bibitem[Hawley et~al., 2014]{hawley2014kepler}
Hawley, S.~L., Davenport, J.~R., Kowalski, A.~F., Wisniewski, J.~P., Hebb, L.,
  Deitrick, R., and Hilton, E.~J. (2014).
\newblock Kepler flares. i. active and inactive m dwarfs.
\newblock {\em The Astrophysical Journal}, 797(2):121.

\bibitem[Hawley and Pettersen, 1991]{hawley1991great}
Hawley, S.~L. and Pettersen, B.~R. (1991).
\newblock The great flare of 1985 april 12 on ad leonis.
\newblock {\em The Astrophysical Journal}, 378:725--741.

\bibitem[Hayashi, 1966]{hayashi1966evolution}
Hayashi, C. (1966).
\newblock Evolution of protostars.
\newblock {\em Annual Review of Astronomy and Astrophysics}, 4(1):171--192.

\bibitem[{Hunten} et~al., 1987]{Hunten87}
{Hunten}, D.~M., {Pepin}, R.~O., and {Walker}, J.~C.~G. (1987).
\newblock {Mass fractionation in hydrodynamic escape}.
\newblock {\em Icarus}, 69:532--549.

\bibitem[Kasting, 1988]{Kasting88}
Kasting, J. (1988).
\newblock {Runaway and moist greenhouse atmospheres and the evolution of earth
  and Venus}.
\newblock {\em Icarus}, 74:472--494.

\bibitem[Kasting et~al., 1993]{kasting1993habitable}
Kasting, J.~F., Whitmire, D.~P., and Reynolds, R.~T. (1993).
\newblock Habitable zones around main sequence stars.
\newblock {\em Icarus}, 101(1):108--128.

\bibitem[Kopparapu, 2013]{kopparapu2013revised}
Kopparapu, R.~K. (2013).
\newblock A revised estimate of the occurrence rate of terrestrial planets in
  the habitable zones around kepler m-dwarfs.
\newblock {\em The Astrophysical Journal Letters}, 767(1):L8.

\bibitem[{Kopparapu} et~al., 2013]{Kopparapu13}
{Kopparapu}, R.~K., {Ramirez}, R., {Kasting}, J.~F., {Eymet}, V., {Robinson},
  T.~D., {Mahadevan}, S., {Terrien}, R.~C., {Domagal-Goldman}, S., {Meadows},
  V., and {Deshpande}, R. (2013).
\newblock {Habitable Zones around Main-sequence Stars: New Estimates}.
\newblock {\em \apj}, 765:131.

\bibitem[Kopparapu et~al., 2014]{kopparapu2014habitable}
Kopparapu, R.~K., Ramirez, R.~M., SchottelKotte, J., Kasting, J.~F.,
  Domagal-Goldman, S., and Eymet, V. (2014).
\newblock Habitable zones around main-sequence stars: dependence on planetary
  mass.
\newblock {\em The Astrophysical Journal Letters}, 787(2):L29.

\bibitem[Lacy et~al., 1976]{lacy1976uv}
Lacy, C.~H., Moffett, T.~J., and Evans, D.~S. (1976).
\newblock Uv ceti stars-statistical analysis of observational data.
\newblock {\em The Astrophysical Journal Supplement Series}, 30:85--96.

\bibitem[Laughlin et~al., 1997]{laughlin1997end}
Laughlin, G., Bodenheimer, P., and Adams, F.~C. (1997).
\newblock The end of the main sequence.
\newblock {\em The Astrophysical Journal}, 482:420--432.

\bibitem[{Lin} and {Ida}, 1997]{LinIda97}
{Lin}, D.~N.~C. and {Ida}, S. (1997).
\newblock {On the Origin of Massive Eccentric Planets}.
\newblock {\em \apj}, 477:781--791.

\bibitem[Linsky et~al., 2013]{linsky2013}
Linsky, J.~L., Fontenla, J., and France, K. (2013).
\newblock {THE} {INTRINSIC} {EXTREME} {ULTRAVIOLET} {FLUXES} {OF} f5 v {TO} m5
  v {STARS}.
\newblock {\em The Astrophysical Journal}, 780(1):61.

\bibitem[Lopez et~al., 2012]{lopez2012thermal}
Lopez, E.~D., Fortney, J.~J., and Miller, N. (2012).
\newblock How thermal evolution and mass-loss sculpt populations of
  super-earths and sub-neptunes: application to the kepler-11 system and
  beyond.
\newblock {\em The Astrophysical Journal}, 761(1):59.

\bibitem[Luger and Barnes, 2015]{luger2015extreme}
Luger, R. and Barnes, R. (2015).
\newblock Extreme water loss and abiotic o2 buildup on planets throughout the
  habitable zones of m dwarfs.
\newblock {\em Astrobiology}, 15(2):119--143.

\bibitem[Luger et~al., 2015]{luger2015habitable}
Luger, R., Barnes, R., Lopez, E., Fortney, J., Jackson, B., and Meadows, V.
  (2015).
\newblock Habitable evaporated cores: transforming mini-neptunes into
  super-earths in the habitable zones of m dwarfs.
\newblock {\em Astrobiology}, 15(1):57--88.

\bibitem[Murray-Clay et~al., 2009]{murray2009atmospheric}
Murray-Clay, R.~A., Chiang, E.~I., and Murray, N. (2009).
\newblock Atmospheric escape from hot jupiters.
\newblock {\em The Astrophysical Journal}, 693(1):23.

\bibitem[Newton et~al., 2017]{newton2017halpha}
Newton, E.~R., Irwin, J., Charbonneau, D., Berlind, P., Calkins, M.~L., and
  Mink, J. (2017).
\newblock The h$\alpha$ emission of nearby m dwarfs and its relation to stellar
  rotation.
\newblock {\em The Astrophysical Journal}, 834(1):85.

\bibitem[Osten and Wolk, 2015]{osten2015connecting}
Osten, R.~A. and Wolk, S.~J. (2015).
\newblock Connecting flares and transient mass-loss events in magnetically
  active stars.
\newblock {\em The Astrophysical Journal}, 809(1):79.

\bibitem[Owen and Wu, 2016]{owen2016atmospheres}
Owen, J.~E. and Wu, Y. (2016).
\newblock Atmospheres of low-mass planets: the “boil-off”.
\newblock {\em The Astrophysical Journal}, 817(2):107.

\bibitem[{Peacock} et~al., 2020]{Peacock2020}
{Peacock}, S., {Barman}, T., {Shkolnik}, E.~L., {Loyd}, R.~O.~P., {Schneider},
  A.~C., {Pagano}, I., and {Meadows}, V.~S. (2020).
\newblock {HAZMAT VI: The Evolution of Extreme Ultraviolet Radiation Emitted
  from Early M Stars}.
\newblock {\em \apj}, 895(1):5.

\bibitem[Ramirez and Kaltenegger, 2014]{ramirez2014habitable}
Ramirez, R.~M. and Kaltenegger, L. (2014).
\newblock The habitable zones of pre-main-sequence stars.
\newblock {\em The Astrophysical Journal Letters}, 797(2):L25.

\bibitem[{Rasio} and {Ford}, 1996]{RasioFord96}
{Rasio}, F.~A. and {Ford}, E.~B. (1996).
\newblock {Dynamical instabilities and the formation of extrasolar planetary
  systems}.
\newblock {\em Science}, 274:954--956.

\bibitem[Ribas et~al., 2005]{ribas2005evolution}
Ribas, I., Guinan, E.~F., G{\"u}del, M., and Audard, M. (2005).
\newblock Evolution of the solar activity over time and effects on planetary
  atmospheres. i. high-energy irradiances (1-1700 {\aa}).
\newblock {\em The Astrophysical Journal}, 622(1):680.

\bibitem[Rodriguez et~al., 2020]{rodriguez2020first}
Rodriguez, J.~E., Vanderburg, A., Zieba, S., Kreidberg, L., Morley, C.~V.,
  Eastman, J.~D., Kane, S.~R., Spencer, A., Quinn, S.~N., Cloutier, R., et~al.
  (2020).
\newblock The first habitable-zone earth-sized planet from tess. ii. spitzer
  confirms toi-700 d.
\newblock {\em The Astronomical Journal}, 160(3):117.

\bibitem[{Salz} et~al., 2016]{Salz2016}
{Salz}, M., {Schneider}, P.~C., {Czesla}, S., and {Schmitt}, J.~H.~M.~M.
  (2016).
\newblock {Energy-limited escape revised. The transition from strong planetary
  winds to stable thermospheres}.
\newblock {\em \aap}, 585:L2.

\bibitem[{Sanz-Forcada, J.} et~al., 2011]{sanzforcada2011}
{Sanz-Forcada, J.}, {Micela, G.}, {Ribas, I.}, {Pollock, A. M. T.}, {Eiroa,
  C.}, {Velasco, A.}, {Solano, E.}, and {Garc\'{\i}a-\'Alvarez, D.} (2011).
\newblock Estimation of the xuv radiation onto close planets and their
  evaporation.
\newblock {\em A\&A}, 532:A6.

\bibitem[{Schaefer} et~al., 2016]{2016ApJ...829...63S}
{Schaefer}, L., {Wordsworth}, R.~D., {Berta-Thompson}, Z., and {Sasselov}, D.
  (2016).
\newblock {Predictions of the Atmospheric Composition of GJ 1132b}.
\newblock {\em \apj}, 829(2):63.

\bibitem[Shields et~al., 2016]{shields2016habitability}
Shields, A.~L., Ballard, S., and Johnson, J.~A. (2016).
\newblock The habitability of planets orbiting m-dwarf stars.
\newblock {\em Physics Reports}, 663:1--38.

\bibitem[Sotin et~al., 2007]{sotin2007mass}
Sotin, C., Grasset, O., and Mocquet, A. (2007).
\newblock Mass--radius curve for extrasolar earth-like planets and ocean
  planets.
\newblock {\em Icarus}, 191(1):337--351.

\bibitem[Tian and Ida, 2015]{tian2015water}
Tian, F. and Ida, S. (2015).
\newblock Water contents of earth-mass planets around m dwarfs.
\newblock {\em Nature Geoscience}, 8(3):177--180.

\bibitem[Tilley et~al., 2019]{tilley2019modeling}
Tilley, M.~A., Segura, A., Meadows, V., Hawley, S., and Davenport, J. (2019).
\newblock Modeling repeated m dwarf flaring at an earth-like planet in the
  habitable zone: atmospheric effects for an unmagnetized planet.
\newblock {\em Astrobiology}, 19(1):64--86.

\bibitem[Torres et~al., 2017]{torres2017validation}
Torres, G., Kane, S.~R., Rowe, J.~F., Batalha, N.~M., Henze, C.~E., Ciardi,
  D.~R., Barclay, T., Borucki, W.~J., Buchhave, L.~A., Crepp, J.~R., et~al.
  (2017).
\newblock Validation of small kepler transiting planet candidates in or near
  the habitable zone.
\newblock {\em The Astronomical Journal}, 154(6):264.

\bibitem[Tuomi et~al., 2019]{tuomi2019frequency}
Tuomi, M., Jones, H., Butler, R., Arriagada, P., Vogt, S., Burt, J., Laughlin,
  G., Holden, B., Shectman, S., Crane, J., et~al. (2019).
\newblock Frequency of planets orbiting m dwarfs in the solar neighbourhood.
\newblock {\em arXiv preprint arXiv:1906.04644}.

\bibitem[Turbet et~al., 2020]{turbet2020revised}
Turbet, M., Bolmont, E., Ehrenreich, D., Gratier, P., Leconte, J., Selsis, F.,
  Hara, N., and Lovis, C. (2020).
\newblock Revised mass-radius relationships for water-rich rocky planets more
  irradiated than the runaway greenhouse limit.
\newblock {\em Astronomy \& Astrophysics}, 638:A41.

\bibitem[Vanderburg et~al., 2020]{vanderburg2020habitable}
Vanderburg, A., Rowden, P., Bryson, S., Coughlin, J., Batalha, N., Collins,
  K.~A., Latham, D.~W., Mullally, S.~E., Col{\'o}n, K.~D., Henze, C., et~al.
  (2020).
\newblock A habitable-zone earth-sized planet rescued from false positive
  status.
\newblock {\em The Astrophysical Journal Letters}, 893(1):L27.

\bibitem[{Watson} et~al., 1981]{1981Icar...48..150W}
{Watson}, A.~J., {Donahue}, T.~M., and {Walker}, J.~C.~G. (1981).
\newblock {The dynamics of a rapidly escaping atmosphere: Applications to the
  evolution of Earth and Venus}.
\newblock {\em \icarus}, 48(2):150--166.

\bibitem[West et~al., 2008]{west2008constraining}
West, A.~A., Hawley, S.~L., Bochanski, J.~J., Covey, K.~R., Reid, I.~N.,
  Dhital, S., Hilton, E.~J., and Masuda, M. (2008).
\newblock Constraining the age-activity relation for cool stars: the sloan
  digital sky survey data release 5 low-mass star spectroscopic sample.
\newblock {\em The Astronomical Journal}, 135(3):785.

\bibitem[Wordsworth and Pierrehumbert, 2013]{wordsworth2013water}
Wordsworth, R. and Pierrehumbert, R. (2013).
\newblock Water loss from terrestrial planets with co2-rich atmospheres.
\newblock {\em The Astrophysical Journal}, 778(2):154.

\bibitem[Wordsworth et~al., 2018]{wordsworth2018redox}
Wordsworth, R., Schaefer, L., and Fischer, R. (2018).
\newblock Redox evolution via gravitational differentiation on low-mass
  planets: Implications for abiotic oxygen, water loss, and habitability.
\newblock {\em The Astronomical Journal}, 155(5):195.

\bibitem[Youngblood et~al., 2017]{youngblood2017muscles}
Youngblood, A., France, K., Loyd, R.~P., Brown, A., Mason, J.~P., Schneider,
  P.~C., Tilley, M.~A., Berta-Thompson, Z.~K., Buccino, A., Froning, C.~S.,
  et~al. (2017).
\newblock The muscles treasury survey. iv. scaling relations for ultraviolet,
  ca ii k, and energetic particle fluxes from m dwarfs.
\newblock {\em The Astrophysical Journal}, 843(1):31.

\end{thebibliography}

\end{document}